\documentclass[12pt]{article}
\usepackage{graphicx}
\begin{document}
\font\bigcal=cmsy10 at 12truept
\def\calE{\hbox{\bigcal E}}
\font\smallcal=cmsy10 at 8truept
\def\scalE{\hbox{\smallcal E}}
\def\bt{{\bar t}\,\strut}
\title{New Results for the Correlation Functions of the Ising Model
and the Transverse Ising Chain}
\author{Jacques H.H. Perk and Helen Au-Yang\thanks{Email:
perk@okstate.edu}\ \thanks{Supported in part by the
National Science Foundation under grant
PHY 07-58139 and by the Australian Research Council under
Project ID: LX0989627}\\\\
{\em\normalsize 145 Physical Sciences, Oklahoma State University,}\\
{\em\normalsize Stillwater, OK 74078-3072,
USA}\thanks{Permanent address}\\\\
{\em\normalsize Department of Theoretical Physics, (RSPE),} {\rm and}\\
{\em\normalsize Centre for Mathematics and its Applications (CMA),}\\
{\em\normalsize Australian National University,}\\
{\em\normalsize Canberra, ACT 2600, Australia}
}
\maketitle

\noindent{\footnotesize
In this paper we show how an infinite system of
coupled Toda-type nonlinear differential equations derived by one of
us can be used efficiently to calculate the time-dependent
pair-correlations in the Ising chain in a transverse field. The results
are seen to match extremely well long large-time asymptotic expansions
newly derived here. For our initial conditions we use new long
asymptotic expansions for the equal-time pair correlation functions
of the transverse Ising chain, extending an old result of T.T.\ Wu for
the 2d Ising model. Using this one can also study the equal-time
wavevector-dependent correlation function of the quantum chain,
a.k.a.\ the $q$-dependent diagonal susceptibility in the 2d Ising model,
in great detail with very little computational effort.}

\section{Introduction}
\label{intro}

In recent years there has been much interest in the properties of
low-dimen\-sional quantum systems. It seems to be worthwhile,
therefore, to present a new algorithm to efficiently and
accurately calculate a very large number of time-dependent
pair correlations in the bulk for the premier example of
such systems, namely for the transverse Ising chain.\footnote{The
transverse Ising chain \cite{Pf} is a special case of the XY model first
introduced by Nambu \cite{Na} in 1950 for the isotropic zero-field
case and generalized by Lieb, Schultz and Mattis \cite{LSM} and
Katsura \cite{Ka}.} Many results have been derived before for this
model,\footnote{We shall not consider the time-dependent $zz$
correlations \cite{Ni,KHS,TH,HT,PCS}, which being only 2-by-2
determinants do not involve ``Jordan--Wigner strings" of fermion
operators.} but the foremost method is still to calculate very large
determinants for each data point, using either the appropriate
determinant for the open chain with both spins far from the
boundary \cite{PM}, or the infinite McCoy--Barouch--Abraham
determinant \cite{MBA} for the closed chain.

In this paper we shall use a set of coupled nonlinear
differential-difference equations derived by one of us \cite{P1,PCQN},
related to similar identities for the planar Ising model
\cite{MW2,P2}. From these we can obtain multiple
millions of data points in one single run. In spite of what some
colleagues have told us---this work is partly an answer to
their unbelief---our nonlinear equations are highly effective for
numerical computations as we shall demonstrate. We
shall concentrate on the zero-temperature case,\footnote{The case
of finite temperature should work out fine also according to
a calculation we have done with approximate initial conditions,
utilizing the exponential decay with separation in the initial
conditions. The case of infinite temperature has been studied
before in great detail \cite{SJL,BJ1,CP,BJ2,PC1,PC2,PC3,SVM}.
We shall use some of the older results for zero temperature
\cite{MBA,JM,VT,MPS1,MPS2,MS1,MS2,MS3,MS4}. Finally, there are several
partial results for finite temperature worth mentioning, e.g.\
\cite{IIKS1,IIKN,CIKT,IIKS2,IIKS3,DZ,SNM,Sa1,Sa2,DG}.} for which the
initial conditions coincide with the diagonal correlations in the
two-dimensional Ising model.

Therefore, we shall first discuss in section 2 how to accurately
calculate these diagonal correlations. For smaller separations we can
use the algorithms of Jimbo and Miwa \cite{JM17} or of Witte \cite{Wi}.
For larger separations we need several more terms in the old
asymptotic expansion of T.T.~Wu \cite{Wu,MW1}, which we shall derive as a
new result.

In section 3 we introduce the transverse-field Ising chain.
We shall also present new results for the asymptotic expansions
for its correlations as a function of time. In section 4 we
shall give details of how we solved the correlations numerically
and show how well the numerical results agree with the
asymptotic expansions derived in section 3.
We conclude with a few remarks in section 5.

\section{Results for the Two-Dimensional Ising\hfill\break Model}
\label{sec2}

In this section we shall present results for the diagonal
pair correlations of the two-dimensional Ising model on the
square lattice and its dual model. These models are described by
the reduced interaction energies (with factor $\beta=1/k_{\mathrm B}T$ absorbed)
\begin{equation}
-\beta\mathcal{E}=\sum_m\sum_n \big(K_1\sigma_{m,n}\sigma_{m+1,n}+
K_2\sigma_{m,n}\sigma_{m,n+1}\big),
\end{equation}
and
\begin{equation}
-(\beta\mathcal{E})^{\,\ast}=\sum_m\sum_n
\big(K_1^{\ast}\sigma_{m,n}\sigma_{m+1,n}+
K_2^{\ast}\sigma_{m,n}\sigma_{m,n+1}\big).
\end{equation}
It is convenient to introduce the following short-hand notations
for the interaction parameters:
\begin{equation}
t_1\equiv\tanh K_1={\mathrm e}^{-2K_2^{\ast}},\quad
t_2\equiv\tanh K_2={\mathrm e}^{-2K_1^{\ast}},
\end{equation}
and
\begin{eqnarray}
S_1\equiv\sinh(2K_1)=1/\sinh(2K_2^{\ast})=1/S_2^{\ast},\cr\cr
S_2\equiv\sinh(2K_2)=1/\sinh(2K_1^{\ast})=1/S_1^{\ast}.
\end{eqnarray}
We shall restrict the elliptic modulus
\begin{equation}
k\equiv S_1S_2
\end{equation}
to $0\le k\le1$, as the case $k>1$ is included via the dual
model,\footnote{The case $k<0$ follows by simple gauge symmetry,
but will not be used.} defining
\begin{equation}
C(m,n)\equiv\langle\sigma_{0,0}\sigma_{m,n}\rangle=
{\sum_{\{\sigma\}}\sigma_{0,0}\sigma_{m,n}{\mathrm e}^{-\beta\mathcal{E}}
\raise-4pt\hbox{\bigg/}
\sum_{\{\sigma\}}{\mathrm e}^{-\beta\mathcal{E}}},
\end{equation}
\begin{equation}
C^{\ast}(m,n)\equiv\langle\sigma_{0,0}\sigma_{m,n}\rangle^{\ast}=
{\sum_{\{\sigma\}}\sigma_{0,0}\sigma_{m,n}{\mathrm e}^{-\beta\mathcal{E}^{\ast}}
\raise-4pt\hbox{\bigg/}
\sum_{\{\sigma\}}{\mathrm e}^{-\beta\mathcal{E}^{\ast}}}.
\end{equation}
As already said before,
for the initial conditions of the quantum chain, we need to calculate
$C(m,n)$ and $C^{\ast}(m,n)$ for $m=n$. This can be done using the
quadratic recurrence relations of Jimbo and Miwa \cite{JM17} or Witte
\cite{Wi}

When $T\ne T_{\mathrm c}$, we can use the well-known Toeplitz determinants
\cite{MW1,AP4}. It is, however, more efficient to use the equations
provided by Jimbo and Miwa \cite{JM17}. Denoting $C^{\ast}(n,n)\equiv A_n$
and $C(n,n)\equiv C_n$, these can be rewritten as:
\begin{eqnarray}
&&B^{\phantom{-}}_{n+1}=
-(kA^+_nB^+_n+k^{-1}A^-_nB^-_n)/
\big((2n+3)A^{\phantom{-}}_{n}\big),
\label{diag1}\\
&&C^{\pm}_{n+1}=
(A^{\phantom{-}}_{n+1}C^{\pm}_n
-C^{\phantom{-}}_nA^{\pm}_n)/
(k^{\pm1}A^{\phantom{-}}_{n}),
\label{diag2}\\
&&D^{\pm}_{n+1}=
(A^{\phantom{-}}_{n+1}D^{\pm}_n
+C^{\phantom{-}}_nB^{\pm}_n)/A^{\phantom{-}}_{n},
\label{diag3}\\
&&C^{\phantom{-}}_{n+1}=
-(C^+_{n+1}D^+_{n+1}+C^-_{n+1}D^-_{n+1})/
\big((2n+1)A^{\phantom{-}}_{n}\big),
\label{diag4}\\
&&A^{\pm}_{n+1}=
(A^{\phantom{-}}_{n+1}A^{\pm}_n
-B^{\phantom{-}}_{n+1}C^{\pm}_{n+1})/A^{\phantom{-}}_{n},
\label{diag5}\\
&&B^{\pm}_{n+1}=
(k^{\pm1}A^{\phantom{-}}_{n+1}B^{\pm}_n
+B^{\phantom{-}}_{n+1}D^{\pm}_{n+1})/A^{\phantom{-}}_{n},
\label{diag6}\\
&&A^{\phantom{-}}_{n+2}=
(A^{\;\;2}_{n+1}-B^{\phantom{-}}_{n+1}C^{\phantom{-}}_{n+1})/
A^{\phantom{-}}_{n},
\label{diag7}
\end{eqnarray}
which can be solved iteratively, in the above order for $n=0,1,\cdots$,
starting from the initial conditions
\begin{eqnarray}
&&A^{\phantom{-}}_0=B^{\phantom{-}}_0=C^{\phantom{-}}_0=1,\quad
A^{\phantom{-}}_1=2{\mathrm E}(k)/\pi,\nonumber\\
&&B^+_0=D^-_0=k',\quad C^+_0=A^-_0={1/k'},\quad
D^+_0=C^-_0=0,
\label{diag8}\\
&&A^+_0=2\big(2\,{\mathrm E}(k)-{\mathrm K}(k)\big)/(\pi k'),\quad
B^-_0=2k'\big({\mathrm K}(k)-{\mathrm E}(k)\big)/\pi,\nonumber
\end{eqnarray}
where K($k$) and E($k$) are the usual complete elliptic integrals,
$k'=\sqrt{1-k^2}$. From these equations one can calculate systematically
more $A_n$'s and $C_n$'s by iteration, keeping sufficiently
many digits, until the asymptotic regime is reached.

The required asymptotic expansions can be obtained from the
Painlev\'e VI equation (PVI) of Jimbo and Miwa \cite{JM17},
extending the expansions of Wu \cite{Wu}---or also eqs.\ (2.46)
and (3.27) of Chapter XI of \cite{MW1}---beyond second order. For solving
PVI iteratively, we need information from the leading terms in the
low- and high-temperature expansions of the connected pair correlations.

The high-temperature expansion of $C(m,n)$ is well-known and is often
treated in graduate courses. Without loss of generality we can take
$m,n\ge0$ and give the leading high-temperature term as
\begin{equation}
C(m,n)\approx\frac{(m+n)!}{m!\,n!}\,t_1^{\,m}t_2^{\,n}
\approx\frac{(m+n)!}{m!\,n!}\,\frac{S_1^{\,m}S_2^{\,n}}{2^{m+n}},
\end{equation}
where the combinatorial factor counts the number of staircase walks
from $(0,0)$ to $(m,n)$ on the square lattice. For the leading term
in the low-tempera\-ture expansion of the connected pair correlation
\begin{equation}
C^{\ast}_{\mathrm c}(m,n)=C^{\ast}(m,n)-(1-k^2)^{1/4},\quad
k\equiv S_1S_2=\frac{1}{S_1^{\ast}S_2^{\ast}},
\end{equation}
subtracting the square of the spontaneous magnetization, we can use
(10a), (10b) or (10c) of \cite{P2}. When doing so we need to replace
$M\to n$, $N\to m$, as currently it is more common to use the horizontal
coordinate as the first one. Rewriting eq.~(10b) of \cite{P2} as
\begin{eqnarray}
&&\Big(C^{\ast}(m-1,n)\,C^{\ast}(m+1,n)-C^{\ast}(m,n)^2\Big)\nonumber\\
&&\qquad+S_2^{\,2}\Big(C(m,n-1)\,C(m,n+1)-C(m,n)^2\Big)=0,
\end{eqnarray}
we can ignore the contributions of $C^{\ast}_{\mathrm c}(m,n)$ and
$C^{\ast}_{\mathrm c}(m+1,n)$ and use $k\to0$. We arrive at
\begin{equation}
C^{\ast}_{\mathrm c}(m-1,n)\approx S_2^{\,2}\Big(C(m,n)^2-C(m,n-1)\,C(m,n+1)\Big),
\end{equation}
or equivalently
\begin{equation}
C^{\ast}_{\mathrm c}(m,n)\approx
S_2^{\,2}\Big(C(m+1,n)^2-C(m+1,n-1)\,C(m+1,n+1)\Big),
\end{equation}
with the leading order solution
\begin{equation}
C^{\ast}_{\mathrm c}(m,n)\approx
\frac{(m+n)!\,(m+n+1)!}{m!\,(m+1)!\,n!\,(n+1)!}\,
\frac{S_1^{\,2m+2}S_2^{\,2n+2}}{2^{2m+2n+2}}.
\label{ghost}
\end{equation}
Here the combinatorial coefficient can be recognized as a
Narayana number, as it counts all staircase polygons needed in the
leading order of the low-temperature expansion.\footnote{We thank
Dr.~Xavier Viennot for pointing this out at the Dunk Island Conference
of 2005. The $q$-analogues of the Narayana numbers were already studied
by MacMahon four decades before Narayana rediscovered them, see
\cite{LM} and references cited therein.} Setting $m=n$, we
get\footnote{These results have been advocated by Dr.~Ranjan Kumar
Ghosh \cite{Gh} as initial conditions for the derivation of the
first few terms in the high- and low-temperature expansions using PVI,
with the second result in (\ref{ghosh}) presented as a conjecture.
However, there is a much more efficient way to derive such
expansions \cite{ONGP1,ONGP2}.}
\begin{equation}
C(n,n)\approx\frac{(2n)!}{(n!)^2}\,\frac{k^{n}}{2^{2n}},\quad
C^{\ast}_{\mathrm c}(n,n)\approx
\frac{(2n)!\,(2n+1)!}{(n!)^2\,[(n+1)!]^2}\,
\frac{k^{2n+2}}{2^{4n+2}}.
\label{ghosh}
\end{equation}

Following Jimbo and Miwa, we define
\begin{eqnarray}
&&t\equiv\frac1{k^2}, \quad
\sigma_n\equiv t(t-1)\frac{{\mathrm d}}{{\mathrm d}t}\ln C(n,n)-{\frac14}t,\nonumber\\
&&\qquad\sigma_n^{\ast}\equiv t(t-1)\frac{{\mathrm d}}{{\mathrm d}t}\ln C^{\ast}(n,n)-{\frac14},
\end{eqnarray}
and then $\sigma_n$ and $\sigma_n^{\ast}$ both satisfy the PVI equation
\begin{eqnarray}
&&\Big[t(t-1)\frac{{\mathrm d}^2\sigma_n}{{\mathrm d}t^2}\Big]^2
-n^2\Big[(t-1)\frac{{\mathrm d}\sigma_n}{{\mathrm d}t}-\sigma_n\Big]^2\nonumber\\
&&\qquad+4\frac{{\mathrm d}\sigma_n}{{\mathrm d}t}
\Big[(t-1)\frac{{\mathrm d}\sigma_n}{{\mathrm d}t}-\sigma_n-{\frac14}\Big]
\Big[t\frac{{\mathrm d}\sigma_n}{{\mathrm d}t}-\sigma_n\Big]=0.
\end{eqnarray}
Substituting
\begin{equation}
C(n,n)=\frac{t^{-n/2}}{\sqrt{\pi n}(1-1/t)^{1/4}}
\exp\bigg(\sum_{j=1}^m\sum_{s=0}^{\lfloor j/2\rfloor}
\frac{p_{j,s}x^{j-2s}}{n^j}\bigg)
\end{equation}
for increasing values of for $m=1,2,\ldots$, while defining
\begin{equation}
x=\frac{t+1}{t-1}=\frac{1+k^2}{1-k^2}
\end{equation}
($x=x_3'$ on page 256 of \cite{MW1}), we can easily solve all the $p_{j,s}$'s.
Whenever $m$ is even, we also need (\ref{ghosh}) in the limit $k\to0$ or
$x\to1$ in order to extract $p_{m,m/2}$.
Similarly, substituting
\begin{equation}
C_{\mathrm c}^{\ast}(n,n)=\frac{t^{-n-1}}{2\pi n^2 (1-1/t)^2}
\exp\bigg(\sum_{j=1}^m\sum_{s=0}^{\lfloor j/2\rfloor}
\frac{p^{\ast}_{j,s}x^{j-2s}}{n^j}\bigg)
\end{equation}
for increasing values of $m=1,2,\ldots$, with the same $x$
($x=-x_3'$ in \cite{MW1} now), we obtain the $p^{\ast}_{j,s}$'s
using (\ref{ghosh}) in addition for the $p^{\ast}_{m,m/2}$ whenever $m$ is
even. Using short Maple programs we went up to $m=20$ and ended with
\begin{equation}
p^{\phantom{\ast}}_{20,10}=\frac{74074237647505}{8388608},\quad
p^{\ast}_{20,10}=\frac{673835095036826977}{10485760}.
\end{equation}
We have used all these results, but here we only give the results
up to $m=10$, i.e.
\begin{eqnarray}
&&C(n,n)=\frac{k^n}{\sqrt{\pi n}(1-k^2)^{1/4}}
\exp\bigg(-\frac{x}{8n}+\frac{(x^2-1)}{16n^2}-\frac{x(25x^2-27)}{384n^3}\nonumber\\
&&\qquad+\,\frac{(x^2-1)(13x^2-5)}{128n^4}-\frac{x(1073x^4-1830x^2+765)}{5120n^5}\nonumber\\
&&\qquad+\,\frac{(x^2-1)(412x^4-425x^2+61)}{768n^6}\nonumber\\
&&\qquad-\,\frac{x(375733x^6-886725 x^4+660723x^2-150003)}{229376n^7}\nonumber\\
&&\qquad+\,\frac{(x^2-1)(23797x^6-40211x^4+18055x^2-1385)}
{4096n^8}\nonumber\\
&&\qquad-\,\frac{\displaystyle{{x(55384775x^8-167281524x^6+179965314x^4\hspace{0.6in}}
\atop{\hspace{1.7in}-79479684x^2
 +11415087)}}}{2359296n^9}\nonumber\\
&&\qquad+\,\frac{\displaystyle{{(x^2-1)(2180461x^8-5127404x^6+3945946x^4\hspace{0.2in}}
\atop{\hspace{1.6in}-1048244x^2
 +50521)}}}{20480n^{10}}-\cdots\bigg)
\label{asyTg}
\end{eqnarray}
and
\begin{eqnarray}
&&C_{\mathrm c}^{\ast}(n,n)=\frac{k^{2n+2}}{2\pi n^2 (1-k^2)^2}
\exp\bigg(-\frac{7x}{4n}+\frac{17x^2-10}{8n^2}-\frac{901x^3-783x}{192n^3}\nonumber\\
&&\qquad+\,\frac{899x^4-1062x^2+194}{64n^4}-\frac{131411x^5-196770x^3+66375x}{2560n^5}\nonumber\\
&&\qquad+\,\frac{83591x^6-151767x^4+75033x^2-6730}{384n^6}\nonumber\\
&&\qquad-\,\frac{17052139x^7-36416187x^5+23770797x^3-4402125x}{16384n^7}\nonumber\\
&&\qquad+\,\frac{11282939x^8-27723492x^6+22515930x^4-6419700x^2+344834}{2048n^8}\nonumber\\
&&\qquad-\,\frac{\displaystyle{{37620804281x^9-104587369452x^7+101707083486x^5
\hspace{0.6in}}
\atop{\hspace{1.8in}-39418182684x^3+4677930225x}}}{1179648n^9}\nonumber\\
&&\qquad+\,\frac{\displaystyle{{2049064082x^{10}-6360721245x^8+7210080180x^6
\hspace{0.1in}}
\atop{\hspace{0.4in}-3544939170x^4
 +670637250x^2-24119050}}}{10240n^{10}}-\cdots\bigg).
\label{asyTs}
\end{eqnarray}

At this point we have an effective way of calculating the connected
diagonal correlation functions of the square-lattice Ising model
and their logarithms, by iteration up to a certain distance using
many-digit precision as this procedure is unstable, and by asymptotic
expansion for larger distances. This way we also have the zero-time
zero-temperature $xx$-correlations of the transverse Ising chain
for $0<k<1$. As we will need very large numbers of $\log C(n,n)$
and $\log C^{\ast}_{\mathrm c}(n,n)$ at fixed $k$ we will have to do the
sums over $s$ only once saving computation time.

For the self-dual case $k=1$, with
$C^{\ast}(n,n)=C^{\ast}_{\mathrm c}(n,n)=C(n,n)$ there is a simple
formula, which was already known to Onsager \cite{FB}, discussed
in detail by McCoy and Wu \cite{MW1}, and implemented by us as:
\begin{equation}
\log C(0,0)=0,\quad \log C(n,n)=\log C(n\!-\!1,n\!-\!1)+r_n,
\end{equation}
\begin{equation}
r_1=\log(2/\pi),\quad r_{n+1}=r_n-\log\big(1-1/(4n^2)\big).
\end{equation}
For this case we also have the large-$n$ asymptotic expansion
\cite{Wu,MW1,AP4}, but we did not need it for this paper.

When $k$ is close to 1, the results are described by 
Painlev\'e III or V (PIII or PV) scaling limit results
\cite{JM17,AP4,WMTB}. Vaidya and Tracy \cite{VT} have analyzed
the time-dependent pair correlations of the transverse Ising
chain in this scaling limit. McCoy, Perk and Shrock \cite{MPS1,MPS2}
found the Painlev\'e II scaling limit describing the crossover
from the space-like to the time-like regime for this quantum chain model.

\section{Quantum Ising Chain}
\label{sec3}

From now on we shall consider equilibrium bulk properties of the
ferromagnetic Ising chain in a positive transverse field. The sites
are labeled by integers $j$, (with $-\mathcal{N}\le j\le\mathcal{N}$, and
$\mathcal{N}\to\infty$ in the thermodynamic limit). We
shall take the usual spin-$\frac12$ operator basis for the quantum chain
\begin{eqnarray}
\sigma^x_j\equiv\cdots\pmatrix{1&0\cr0&1\cr}\otimes\pmatrix{1&0\cr0&1\cr}
\otimes&\overbrace{
\pmatrix{0&\,1\,\cr1&0\cr}}^{\hbox{$j$-th}}&\otimes\pmatrix{1&0\cr0&1\cr}
\otimes\pmatrix{1&0\cr0&1\cr}\otimes\cdots,\nonumber\\\nonumber\\
\sigma^y_j\equiv\cdots\pmatrix{1&0\cr0&1\cr}\otimes\pmatrix{1&0\cr0&1\cr}
\otimes&\pmatrix{0&-{\mathrm i}\cr{\mathrm i}&0\cr}&\otimes\pmatrix{1&0\cr0&1\cr}
\otimes\pmatrix{1&0\cr0&1\cr}\otimes\cdots,\nonumber\\\nonumber\\
\sigma^z_j\equiv\cdots\pmatrix{1&0\cr0&1\cr}\otimes\pmatrix{1&0\cr0&1\cr}
\otimes&\pmatrix{1&0\cr0&-1\cr}&\otimes\pmatrix{1&0\cr0&1\cr}
\otimes\pmatrix{1&0\cr0&1\cr}\otimes\cdots.\nonumber\\
\end{eqnarray}
The spin operators obey the usual Schr\"odinger time-dependence
in the\break Heisenberg picture
\begin{equation}
\sigma^{\alpha}_j(t)\equiv{\mathrm e}^{{\mathrm i}{\mathcal H}t}
\sigma^{\alpha}_j\,{\mathrm e}^{-{\mathrm i}{\mathcal H}t}\,\quad
\hbox{(in units for which $\hbar\equiv1$,
$\alpha=x,y,z$)},
\end{equation}
with Hamiltonian $\mathcal H$ (or dual Hamiltonian $\mathcal H^{\ast}$)
\begin{equation}
{\mathcal H}=-{\textstyle{\frac12}}\sum_{j=-\infty}^{\infty}
(J\sigma_j^x\sigma_{j+1}^x+B\sigma_j^z), \quad
{\mathcal H}^{\ast}=-{\textstyle{\frac12}}\sum_{j=-\infty}^{\infty}
(B\sigma_j^x\sigma_{j+1}^x+J\sigma_j^z),
\label{tiham}
\end{equation}
with $J,B>0$. The dual chain corresponds to the interchange of
$J$ and $B$.\footnote{Strictly spoken, the dual positions are
at the half-integers, but we can relabel them to the integers.
Also, as the bulk thermodynamic limit for pair correlations
has been well-established and is independent
of boundary conditions, we have formally replaced $\mathcal N$ by
$\infty$ in the bounds of the summations in (\ref{tiham}). This
thermodynamic limit is to be understood to be part of the
pair correlation's definition in the following.}
The pair correlation function
\begin{equation}
X_n(t)\equiv\langle\sigma_j^x(t)\sigma_{j+n}^x\rangle\equiv
\frac{\mathrm{Tr}\,({\mathrm e}^{{\mathrm i}t\mathcal H}\sigma_j^x\,
{\mathrm e}^{-{\rm i}t\mathcal H}
\sigma_{j+n}^x\,{\mathrm e}^{-\beta\mathcal H})}
{\mathrm{Tr}\,({\mathrm e}^{-\beta\mathcal H})}
\label{TIcorr}
\end{equation}
and the dual $X_n^{\ast}(t)$, with $\mathcal H$ replaced by
$\mathcal H^{\ast}$, satisfy \cite{P1,PCQN}
\begin{equation}
\left\{\matrix{X_n(t)\ddot X_n(t)-\dot X_n(t)^2=
B^2\Big(X_{n-1}^{\ast}(t)X_{n+1}^{\ast}(t)-X_n^{\ast}(t)^2\Big)\,,\cr\cr
X_n^{\ast}(t)\ddot X_n^{\ast}(t)-\dot X_n^{\ast}(t)^2=
J^2\Big(X_{n-1}(t)X_{n+1}(t)-X_n(t)^2\Big)\,,}\right.
\label{dToda}
\end{equation}
for all inverse temperatures $\beta=1/k_{\mathrm B}T$.
At the critical field $B=J$ this reduces to \cite{P1,PCQN}
\begin{equation}
X_n(t)\ddot X_n(t)-\dot X_n(t)^2=
J^2\big(X_{n-1}(t)X_{n+1}(t)-X_n(t)^2\big).
\label{Toda}
\end{equation}

For real times $t$ these equations represent a discrete nonlinear generalization of a system of hyperbolic partial differential
equations. Therefore, we expect the initial-value problem to be
stable and this is indeed the case for sufficiently small time steps.
On the other hand, for imaginary times $t$, as needed for the
calculation of the susceptibility, we have an elliptic system
and we should treat these equations as a boundary value problem
or work with an increasingly high number of digits as $|t|$ increases,
as was done before for the (euclidian) two-dimensional Ising model,
first for the uniform case \cite{KAP1,KAP2,KAP3,Ko} and later for
quasiperiodic cases \cite{AJP,AP3,AP7,AP8}.\footnote{These latter
cases could be done because the general quadratic identities
of \cite{P1} apply also to $Z$-invariant cases \cite{Bax}
with further implications for correlations and
susceptibilities \cite{AP4,AP2,AP5,AP6}.}
In fact, the susceptibility for the transverse Ising model has been
calculated in the mid 1980s as an extremely anisotropic limit of
the corresponding 2d Ising result \cite{KAP1,KAP2,KAP3,Ko}.

Before describing the algorithm used, we shall first derive
some long-time asymptotic expansions, extending the results
on the last page of \cite{MBA}, while also implicitly correcting several
misprints there. We start with adopting the results (2.1) and (2.7) in
\cite{VT} to our case.\footnote{There are misprints in
(1.2) and (1.3) of \cite{VT}, which differ from the actual definitions
used in \cite{MBA,CP,PC1,PC3,VT}. The difference means
that the results change to their complex conjugates, or equivalently
to the replacement $t\to-t$.} For $T=0$,
\begin{eqnarray}
X_R(t)&=&\bigg(1-\frac BJ\bigg)^{1/4}
\exp\Big(-\sum_{n=1}^\infty F^{(2n)}\Big),\quad (0<B<J),
\label{XRs}\\
&=&\bigg(1-\frac JB\bigg)^{1/4} X^{(2n-1)}
\exp\Big(-\sum_{n=1}^\infty F^{(2n)}\Big),\quad (0<J<B),
\label{XRg}
\end{eqnarray}
where
\begin{eqnarray}
&&F_>^{(2n)}(R,t)=F_<^{(2n)}(R,t)=F^{(2n)}(R,t)\nonumber\\\cr
&&\hspace*{2.0cm}
=\frac{1}{2n}\int_{-\pi}^{\pi}\mathrm{d}\phi_{1}\cdots
\int_{-\pi}^{\pi}\mathrm{d}\phi_{2n}
\prod_{j=1}^{2n}\Big(L(\phi_j)M(\phi_j,\phi_{j+1})\Big),
\label{fVT}
\end{eqnarray}
and
\begin{eqnarray}
&&X^{(2n-1)}(R,t)=
\int_{-\pi}^{\pi}\mathrm{d}\phi_{1}\cdots
\int_{-\pi}^{\pi}\mathrm{d}\phi_{2n-1}\nonumber\\\cr
&&\hspace*{1.0cm}\times\quad\Bigg[\prod_{j=1}^{2n-2}
\Big(L(\phi_j)M(\phi_j,\phi_{j+1})\Big)\Bigg]
L(\phi_{2n-1})\bar M(\phi_{2n-1},\phi_1).
\label{xVT}
\end{eqnarray}
Here,
\begin{eqnarray}
&&L(\phi)\equiv\frac{\exp\Big(-\mathrm{i}R\phi-\mathrm{i}t\lambda(\phi)\Big)}{4\pi\lambda(\phi)},\\
&&M(\phi,\phi')\equiv
\frac{\lambda(\phi)-\lambda(\phi')}{\sin\Big(\frac12(\phi+\phi')\Big)}=
\frac{4JB\sin\Big(\frac12(\phi-\phi')\Big)}{\lambda(\phi)+\lambda(\phi')},\\\cr
&&\bar M(\phi,\phi')\equiv 2B\cos\Big(\textstyle{\frac12}(\phi-\phi')\Big),\\\cr
&&\lambda(\phi)\equiv\sqrt{J^2+B^2-2JB\cos\phi}.
\end{eqnarray}
It is easily checked that the integrands in (\ref{fVT}) and 
(\ref{xVT}) are periodic modulo $2\pi$ in all the $\phi_j$. Therefore,
before applying the stationary phase method for large $t$, we can
shift the integration bounds and consider the stationary phase
points $\phi_j=0$ and $\phi_j=\pi\equiv-\pi$ as internal point for
the integration over $\phi_j$. Expanding the integrands of
(\ref{fVT}) and (\ref{xVT}) in power series both in $\phi_j$ and in
$\phi_j\!-\!\pi$ for all $j$, keeping only the quadratic terms in the
exponentials of the $L(\phi_j)$'s, we can then use
\begin{equation}
\int_{-\infty}^{+\infty}\mathrm{d}\phi\,
\mathrm{e}^{\pm\mathrm{i}\alpha\phi^2}\phi^{2n-1}=0,\quad
\int_{-\infty}^{+\infty}\mathrm{d}\phi\,
\mathrm{e}^{\pm\mathrm{i}\alpha\phi^2}\phi^{2n}=
\mathrm{e}^{\pm(2n+1)\mathrm{i}\pi/4}\,
\frac{\Gamma(n\!+\!\frac12)}{\alpha^{n+\frac12}}.
\end{equation}
for $\alpha>0$. The leading terms come from $X^{(1)}$ and $F^{(2)}$
and to leading order one finds
\begin{equation}
X^{(2n+1)}\approx \Big(F^{(2)}\Big)^n X^{(1)},\quad
F^{(2n)}\approx\frac1n\Big(F^{(2)}\Big)^n,
\end{equation}
as the stationary phase points contributing are alternatingly
0 and $\pi$, starting with $\phi_1=0$ or $\pi$. In higher orders of
$1/t$ more and more non-alternating combinations come in and it soon
becomes very tedious and unpresentable. We shall, therefore, only
give the results for $R=0$ in (\ref{XRs}) and (\ref{XRg}) here.

We use the abbreviations
\begin{equation}
k=\frac BJ,\quad \bar t=Jt\quad(\mbox{for }B<J),\qquad
k=\frac JB,\quad \bar t=Bt\quad(\mbox{for }B>J).
\end{equation}
Then, for $B<J$ and $t\to\infty$,
\begin{eqnarray}
X_0(t)&=&\left(1-k^2\right)^{1/4}\left(1
+\,\frac{k\,{\rm e}^{-2i\bar t}}
  {2\pi\,\sqrt{1-k^2}\,\bar t}
  \right.\nonumber\\\cr
&&+\,\frac{ik(4-3k^2){\rm e}^{-2i\bar t}}%
  {8\pi\left(1-k^2\right)^{3/2}\bt^{\,2}}
-\frac{{\rm e}^{-2i(1-k)\bar t}}
  {8\pi\left(1-k\right)^2\bt^{\,2}}
-\frac{{\rm e}^{-2i(1+k)\,\bar t}}
  {8\pi\left(1+k\right)^2\bt^{\,2}}\nonumber\\\cr
&&+\,\frac{(4-60k^2+80k^4-33k^6){\rm e}^{-2i\bar t}}
  {64\pi k\left(1-k^2\right)^{5/2}\bt^{\,3}}\nonumber\\\cr
&&+\frac{i(1-9k+k^2){\rm e}^{-2i(1-k)\bar t}}
  {32\pi k\left(1-k\right)^3\bt^{\,3}}
-\frac{i(1+9k+k^2){\rm e}^{-2i(1+k)\bar t}}
  {32\pi k\left(1+k\right)^3\bt^{\,3}}\nonumber\\\cr
&&\left.-\,\frac{ik^2(2+k^2)\,{\rm e}^{-4i\bar t}}
  {32\pi^2\left(1-k^2\right)^2\bt^{\,3}}
+\cdots\right),
\label{asyJgB}
\end{eqnarray}
whereas, for $B>J$ and $t\to\infty$,
\begin{eqnarray}
X_0(t)&=&\left(1-k^2\right)^{1/4}
\left(\,\frac{{\rm e}^{-i(1-k)\bar t-\pi i/4}}
  {\sqrt{2\pi}\left(k(1-k)\bt\,\right)^{1/2}}
+\,\frac{{\rm e}^{-i(1+k)\bar t+\pi i/4}}
  {\sqrt{2\pi}\left(k(1+k)\bt\,\right)^{1/2}}\right.\nonumber\\\cr
&&-\,\frac{\left(1-3k+k^2\right){\rm e}^{-i(1-k)\bar t+\pi i/4}}
  {8\sqrt{2\pi}\left(k(1-k)\bt\,\right)^{3/2}}
-\,\frac{\left(1+3k+k^2\right){\rm e}^{-i(1+k)\bar t-\pi i/4}}
  {8\sqrt{2\pi}\left(k(1+k)\bt\,\right)^{3/2}}\nonumber\\\cr
&&-\,\frac{3\left(3-10k+17k^2-10k^3+3k^4\right)
    {\rm e}^{-i(1-k)\bar t-\pi i/4}}
  {128\sqrt{2\pi}\left(k(1-k)\bt\,\right)^{5/2}}\nonumber\\\cr
&&-\,\frac {3\left( 3+10k+17k^2+10k^3+3k^4\right)
    {\rm e}^{-i(1+k)\bar t+\pi i/4}}
{128\sqrt {2\pi }\left(k(1+k)\bt\,\right)^{5/2}}\nonumber\\\cr
&&-\,\frac{k^{3/2}\,{\rm e}^{-i(3-k)\bar t+\pi i/4}}
  {4\left(2\pi\right)^{3/2}\left(1-k\right)^2
    \left(1+k\right)^{1/2}\,\bt^{\,5/2}}
\nonumber\\\cr
&&\left.-\,\frac{k^{3/2}\,{\rm e}^{-i(3+k)\bar t-\pi i/4}}
  {4\left(2\pi\right)^{3/2}\left(1+k\right)^2
    \left(1-k\right)^{1/2}\,\bt^{\,5/2}}
+\cdots\right).
\label{asyBgJ}
\end{eqnarray}
The lowest order terms that decay as $t^{-1/2}$ and $t^{-1}$ have
been given before in eq.\ (5.14) of \cite{MBA}, and they agree after one
corrects some misprints in \cite{MBA}.

At the critical field $B=J$, we can use the Painlev\'e V equation
of \cite{MPS1} to obtain many terms in the long-time asymptotic expansion
of $X_0(t)$. Writing
\begin{equation}
X_0(t)=\mathrm{e}^{x^2/8}\tau_0(x),\quad
\sigma_0(x)=x\frac{\mathrm{d}\log\tau_0(x)}{\mathrm{d}x},\quad x\equiv2iJt,
\end{equation}
we must solve
\begin{equation}
\Bigg(x\frac{\mathrm{d}^2\sigma}{\mathrm{d}x^2}\Bigg)^2
+4\Bigg(x\frac{\mathrm{d}\sigma}{\mathrm{d}x}-\sigma\Bigg)
\Bigg[x\frac{\mathrm{d}\sigma}{\mathrm{d}x}+
\Bigg(\frac{\mathrm{d}\sigma}{\mathrm{d}x}\Bigg)^2-\sigma\Bigg]=0,
\label{PV}
\end{equation}
requiring the large-$x$ asymptotic behaviors
\begin{equation}
\sigma_0(x)\approx-\frac{x^2}{4}+z\sqrt{x}+
\sum_{n=0}^{N}\sum_{m=0}^{\lfloor\frac12 n+1\rfloor}
c_{n,m}\frac{z^{n+2-2m}}{x^{n/2}},
\label{sigmaN}
\end{equation}
and
\begin{equation}
\log\tau_0(x)\approx-\frac{x^2}{8}+\log\frac{A}{(x/2)^{1/4}}+
\sum_{n=1}^{N}\sum_{m=0}^{\lfloor\frac12 n\rfloor}
d_{n,m}\frac{z^{n-2m}}{x^{n/2}},
\end{equation}
where \cite{MPS1,MT}
\begin{equation}
x\to\infty,\quad
z\equiv-\frac{1}{\sqrt{2\pi}}\,\mathrm{e}^{-x},\quad
A=2^{1/12}\,\mathrm{e}^{3\zeta'(-1)}.
\end{equation}
(Here we use the more common real definition of $A$ without the
$\mathrm{e}^{-i\pi/8}$ factor of \cite{MPS1}. This factor is here
supplied by the $i^{1/4}$ of $x^{1/4}$.)

We can solve the coefficients $c_{n,m}$ recursively from the sets of
linear equations obtained by using the leading term of order
$\mathrm{O}(x^{(5-N)/2})$ when substituting (\ref{sigmaN}) (with
$N=0,1,\dots$, in succession) into (\ref{PV}). After each step in this
process there are two undetermined coefficients left that are determined
in the next two steps. The coefficients $d_{n,m}$ follow recursively for
$n=3,4,\ldots$, from
\begin{equation}
(n-2m)\,d_{n,m}=-c_{n-2,m}-\frac12(n-2)\,d_{n-2,m-1},
\end{equation}
using also
\begin{equation}
d_{n,0}=-\frac1n,\quad d_{2,1}=0,\quad d_{2n,n}=-\frac{c_{2n,n+1}}n.
\end{equation}
The final result, after exponentiating $\log\tau_0(x)$, is
\begin{eqnarray}
X_0(t)&=&\frac{A}{(x/2)^{1/4}}
\Bigg(1-\frac{z}{x^{1/2}}+\frac{9\,z}{8\,x^{3/2}}-
\frac{2\,z^2-1}{8\,x^2}-\frac{297\,z}{2^7\,x^{5/2}}
\nonumber\\\cr
&&\quad+\,\frac{15\,z^2}{16\,x^3}+\frac{7587\,z}{2^{10}\,x^{7/2}}
-\frac{489\,z^2-81}{2^7\,x^4}
+ \frac{1024\,z^3-1027035\,z}{2^{15}\,x^{9/2}}
\nonumber\\\cr
&&\quad+\,\frac{9387\,z^2}{2^9\,x^5}-
\frac{76800\,z^3-43594695\,z}{2^{18}\,x^{11/2}}
\nonumber\\\cr
&&\quad-\,\frac {851427\,z^2-90072}{2^{13}\,x^6}
+\frac{9094144\,z^3-4418168445\,z}{2^{22}\,x^{13/2}}
\nonumber\\\cr
&&\quad+\,\frac{22520925\,z^2}{2^{15}\,x^7}
-\frac{529007616\,z^3-260700970635\,z}{2^{25}\,x^{15/2}}
\nonumber\\\cr
&&\quad+\,\frac{768\,z^4-1368815805\,z^2+108135000}{2^{18}\,x^8}
\nonumber\\\cr
&&\quad+\,\frac{258931316736\,z^3-139999291654995\,z}{2^{31}\,x^{17/2}}
\nonumber\\\cr
&&\quad-\,\frac{59904\,z^4-47100085335\,z^2}
{2^{20}\,x^9}
\nonumber\\\cr
&&\quad-\,\frac{17114655467520\,z^3-10543684346529075\,z}{2^{34}\,x^{19/2}}
\nonumber\\\cr
&&\quad+\,\frac{24846336\,z^4-14491877193315\,z^2+908002224000}{2^{25}\,x^{10}}
\nonumber\\\cr
&&\quad+\,\frac{2469862452602880\,z^3-1758124895330287575\,z}{2^{38}\,x^{21/2}}
\nonumber\\\cr
&&\quad-\,\frac{1113772032\,z^4-616117763829645\,z^2}{2^{27}\,x^{11}}
\nonumber\\\cr
&&\quad-\,
\frac{195069658230835200\,z^3-160841975585736493125\,z}{2^{41}\,x^{23/2}}
\nonumber\\\cr
&&\qquad+\,\mathrm{O}(x^{-12})\Bigg)
\label{asycrit}
\end{eqnarray}
The first line of the above equation agrees with (43) in \cite{MPS1}.
Here we have given a more efficient method to extend the asymptotic
expansion, using real rational coefficients only.%
\footnote{Replacing the term $z\sqrt{x}$ in (\ref{sigmaN}) by 
$-4Kx+(z+K/z)\sqrt{x}$ and summing $m$ from 0 to $n+2$, we can
easily extend also (39) in \cite{MPS1}, correcting two misprints in the
highest order terms there in the process. One must then identify
$K=-iab$ and $z=b\,\mathrm{e}^{-is-i\pi/4}$.}

\section{Numerical Results for Quantum Ising\hfill\break Chain}
\label{sec4}

At zero temperature, the initial values relate to the diagonal correlation
function $\langle\sigma_{00}\sigma_{nn}\rangle$ of the two-dimensional
Ising model discussed in detail in section 2. More precisely,
\begin{equation}
X_n(0)=\langle\sigma_{00}\sigma_{nn}\rangle,\quad
\dot X_n(0)=\dot X_0(0)\,\delta_{n0}=-iB\,
\langle\sigma_0^z\rangle\,\delta_{n0}.
\label{init}
\end{equation}
Here, $k=J/B=S_1S_2<1$ corresponds to the 2d Ising model with
$T>T_{\mathrm c}$, while the case $k=B/J=S_1^{\ast}S_2^{\ast}<1$
corresponds to $T<T_{\mathrm c}$. The initial value of the first
time-derivative $\dot X_n(0)$ vanishes for $n\ne0$ and for $n=0$ it
is given by the $z$-magnetization of the transverse Ising chain
\cite{Pf,Ka,BM},
\begin{eqnarray}
&&\langle\sigma_0^z\rangle=\frac{2}{\pi}\,\mathrm{E}(k),
\phantom{\Big(k-(1-k^2)\mathrm{K}(k)\Big)}
\quad k=\frac{J}{B}\le1,
\label{initJB}\\\cr
&&\langle\sigma_0^z\rangle=\frac{2}{\pi k}\,\Big(\mathrm{E}(k)-
(1-k^2)\,\mathrm{K}(k)\Big),\quad k=\frac{B}{J}\le1,
\label{initBJ}
\end{eqnarray}
where $\mathrm{K}(k)$ and $\mathrm{E}(k)$ are the complete elliptic
integrals of the first and second kind.
For $J=B$, we have the simple results, see e.g.\ \cite{MW1,MPS1},
\begin{equation}X_n(0)=\bigg(\frac{2}{\pi}\bigg)^n\;
\prod_{l=1}^{n-1}\bigg(1-{\frac14 l^2}\bigg)^{l-n},\quad
\dot X_n(0)=-iB\,\frac{2}{\pi}\,\delta_{n0}.
\label{initJJ}
\end{equation}
We have tacitly assumed $n\ge1$ in the above. But, because of
reflection symmetry and reality of the Hamiltonian $\mathcal{H}$,
we have
\begin{equation}
X_{-n}(t)=X_n(t)=X_n(-t)^{\ast},
\end{equation}
where the asterisk denotes complex conjugation.

The results in (\ref{initJJ}) are sufficient by themselves for $B=J$.
But, also for $B\ne J$ we know the initial conditions now to high
precision from earlier work and the newly derived asymptotic
expansions for large $n$ in section 2. From these results we can
find as many time-derivatives at $t=0$ as we want using the differential
equations (\ref{dToda}) and (\ref{Toda}). We can then calculate
$X_n(\delta t)$ and $\dot X_n(\delta t)$ and their duals if $B\ne J$,
to high precision using Taylor expansion to five orders for sufficiently
small $\delta t$, or even more orders. Repeating this process $N$ times
we can calculate $X_n(N\delta t)$ and $\dot X_n(N\delta t)$ (and their
duals) using initial conditions in the ``past light-cone."
This is a discrete version of the method of characteristics. This is
appropriate, since (\ref{dToda}) and (\ref{Toda}) are nonlinear
differential-difference equations of hyperbolic type. In the
continuum limit and for $B=J$, $\log X_n(t)$ satisfies the
two-dimensional Klein--Gordon equation \cite{PCQN}. In the more
general continuum scaling limit $B\to J$ we obtain the hyperbolic
sine-Gordon equation \cite{PCQN,VT}.


\subsection{The case of critical transverse field $B=J$}
\label{sec4a}

When $B=J$, me must solve (\ref{Toda}) with initial conditions
(\ref{initJJ}). It is easily checked that $X_n(t)$ only depends on
$n$ and $\bar t\equiv Jt$. Equivalently we can choose the unit of
time such that $J=1$ and $\bar t\equiv t$, which we shall do in this
subsection.

It is also convenient to work with the new variables
\begin{eqnarray}
\xi_n(t)\equiv\log X_n(t), \quad \dot\xi_n(t)\equiv
\frac{\mathrm{d}\xi_n(t)}{\mathrm{d}t}=\frac{\dot X_n(t)}{X_n(t)},\\
\eta_n(t)\equiv
\frac{X_{n+1}X_{n-1}}{X_n^{\,2}}=\exp(\xi_{n+1}+\xi_{n+1}-2\xi_n).
\end{eqnarray}
Knowing these one can get the higher time-derivatives from (\ref{Toda}), i.e.,
\begin{eqnarray}
\ddot\xi_n(t)&=&\eta_n-1,\\
\stackrel{{\!}_{\textstyle.}{\!}_{\textstyle.}%
{\!}_{\textstyle.}}{\xi_n}\!\!(t)&=&
\eta_n(\dot\xi_{n+1}+\dot\xi_{n-1}-2\dot\xi_{n}),\\
\stackrel{{\!}_{\textstyle.}{\!}_{\textstyle.}%
{\!}_{\textstyle.}{\!}_{\textstyle.}}{\xi_n}\!\!(t)&=&
\eta_n\Big(\ddot\xi_{n+1}+\ddot\xi_{n-1}-2\ddot\xi_{n}+
(\dot\xi_{n+1}+\dot\xi_{n-1}-2\dot\xi_{n})^2\Big),\\
\stackrel{{\!}_{\textstyle.}{\!}_{\textstyle.}{\!}_{\textstyle.}%
{\!}_{\textstyle.}{\!}_{\textstyle.}}{\xi_n}\!\!(t)&=&
\eta_n\Big(\stackrel{{\!}_{\textstyle.}{\!}_{\textstyle.}%
{\!}_{\textstyle.}}{\xi}_{n+1}+
\stackrel{{\!}_{\textstyle.}{\!}_{\textstyle.}%
{\!}_{\textstyle.}}{\xi}_{n-1}-
2\stackrel{{\!}_{\textstyle.}{\!}_{\textstyle.}%
{\!}_{\textstyle.}}{\xi_n}
\nonumber\\
&&\quad+3(\ddot\xi_{n+1}+\ddot\xi_{n-1}-2\ddot\xi_{n})
(\dot\xi_{n+1}+\dot\xi_{n-1}-2\dot\xi_{n})\nonumber\\
&&\quad+(\dot\xi_{n+1}+\dot\xi_{n-1}-2\dot\xi_{n})^3\Big),
\end{eqnarray}
and we can then calculate $\xi_n(t+\delta t)$ and $\dot\xi_n(t+\delta t)$
in Taylor series in $\delta t$ to fifth, respectively fourth order.

\begin{figure}[tbh]
\begin{center}
\includegraphics[width=0.5\hsize]{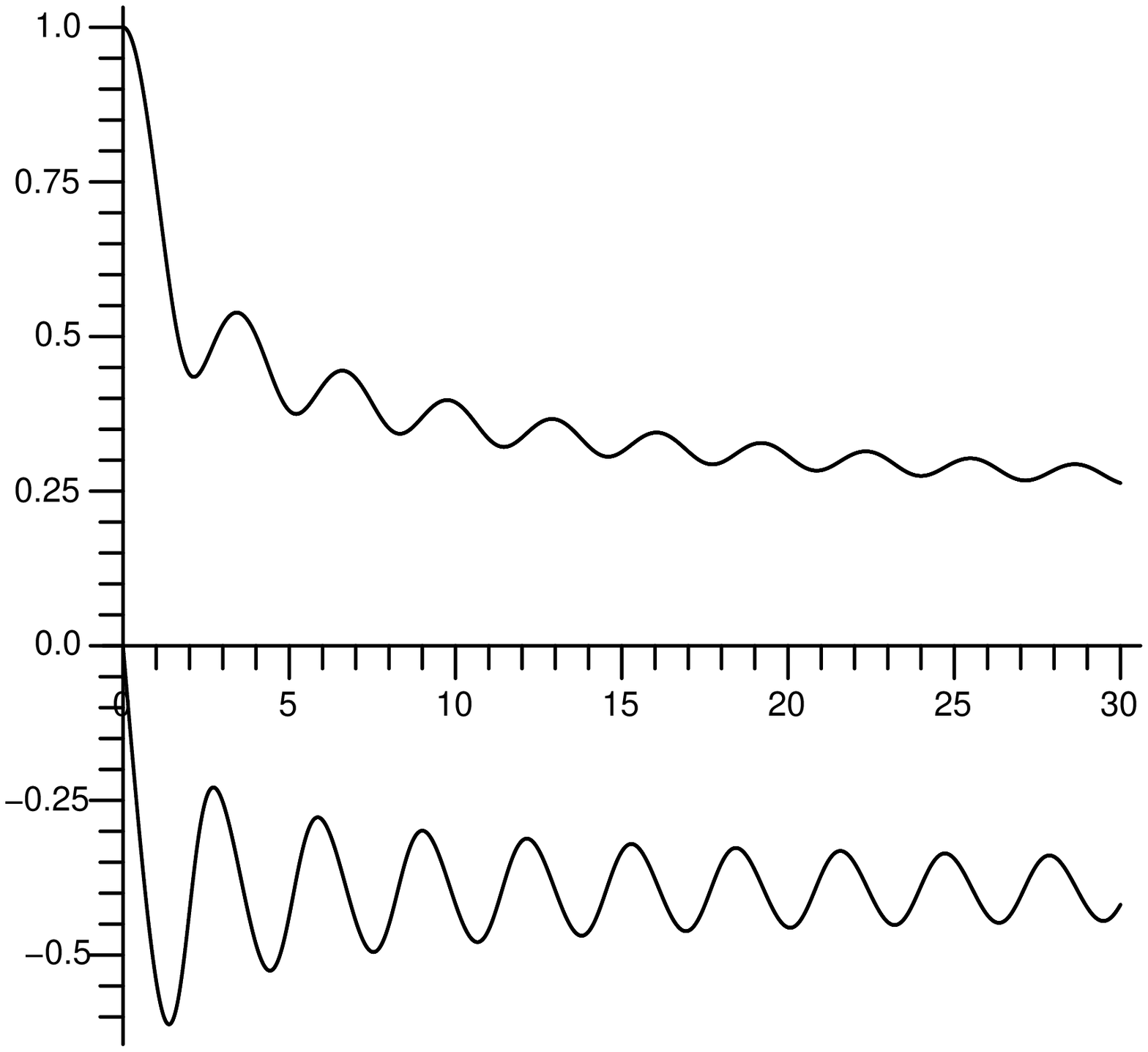}\hfil
\includegraphics[width=0.5\hsize]{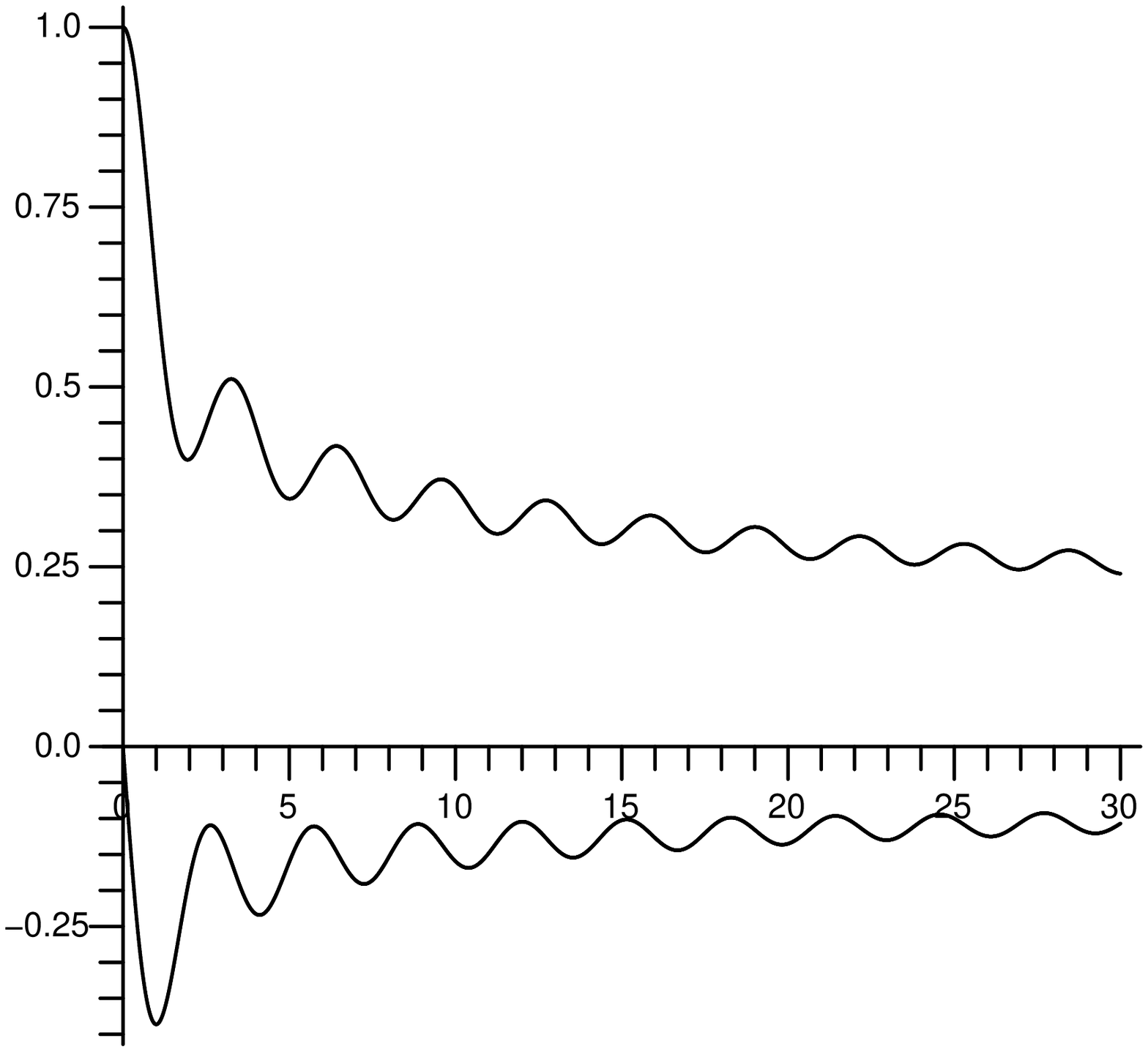}
\end{center}
\caption{$X_0(t)$ for $B=J=1$. On the left are plotted
$|X_0(t)|$ and $\arg X_0(t)$, while on the right $\Re X_0(t)$,
and $\Im X_0(t)$ are shown.}
\end{figure}

We have implemented this on a Macintosh computer using GNU Fortran
with $\delta t=10^{-4}$. Starting at $t=0$ we did $N=300,000$ time steps,
going up to $t=30$. Therefore, we calculated $X_n(t)$ in the triangle
in the $n$-$t$ plane with the three corners $(n,t)$ given by
$(\pm6\!\times\!10^5,0)$ and $(0,30)$. Some of the results are given in
Fig.~1. Comparing the Fortran output for $X_0(t)$ with the asymptotic
expansion (\ref{asycrit}), we concluded that we have already about 12 place
agreement for $t>20$, even though the Fortran program only worked with
15 place floating point complex numbers. This is as good as we could
have hoped for and it demonstrates the stability of the procedure. It is
a strong indication that all the Fortran output within
the ``characteristic triangle" is equally accurate.

Next, we can use the general identities of Lajzerowicz and
Pfeuty \cite{LP},
\begin{eqnarray}
C_n(t)&\equiv&\langle\sigma_j^x(t)\sigma_{j+n}^y\rangle=
-\langle\sigma_j^y(t)\sigma_{j+n}^x\rangle=
\frac{1}{B}\frac{\mathrm{d}X_n(t)}{\mathrm{d}t},\\
Y_n(t)&\equiv&\langle\sigma_j^y(t)\sigma_{j+n}^y\rangle=
-\frac{1}{B^2}\frac{\mathrm{d}^2X_n(t)}{\mathrm{d}t^2},
\label{LazerP}
\end{eqnarray}
which are valid for all $J$ and $B$ and also for finite temperatures.
These quantities are also calculated in the Fortran program with
$B=J=1$ as
\begin{equation}
C_n(t)=\dot\xi_n(t)\,\mathrm{e}^{\xi_n(t)},\quad
Y_n(t)=-\Big(\ddot\xi_n(t)+\dot\xi_n(t)^2\Big)\,\mathrm{e}^{\xi_n(t)}.
\end{equation}
Their asymptotic expansions for $n=0$ follow immediately from
(\ref{asycrit}) and agree with the Fortran data with a similar accuracy
as the $X_0(t)$ did. We have plotted $Y_0(t)$ in Fig.~2.

\begin{figure}[tbh]
\begin{center}
\includegraphics[width=0.5\hsize]{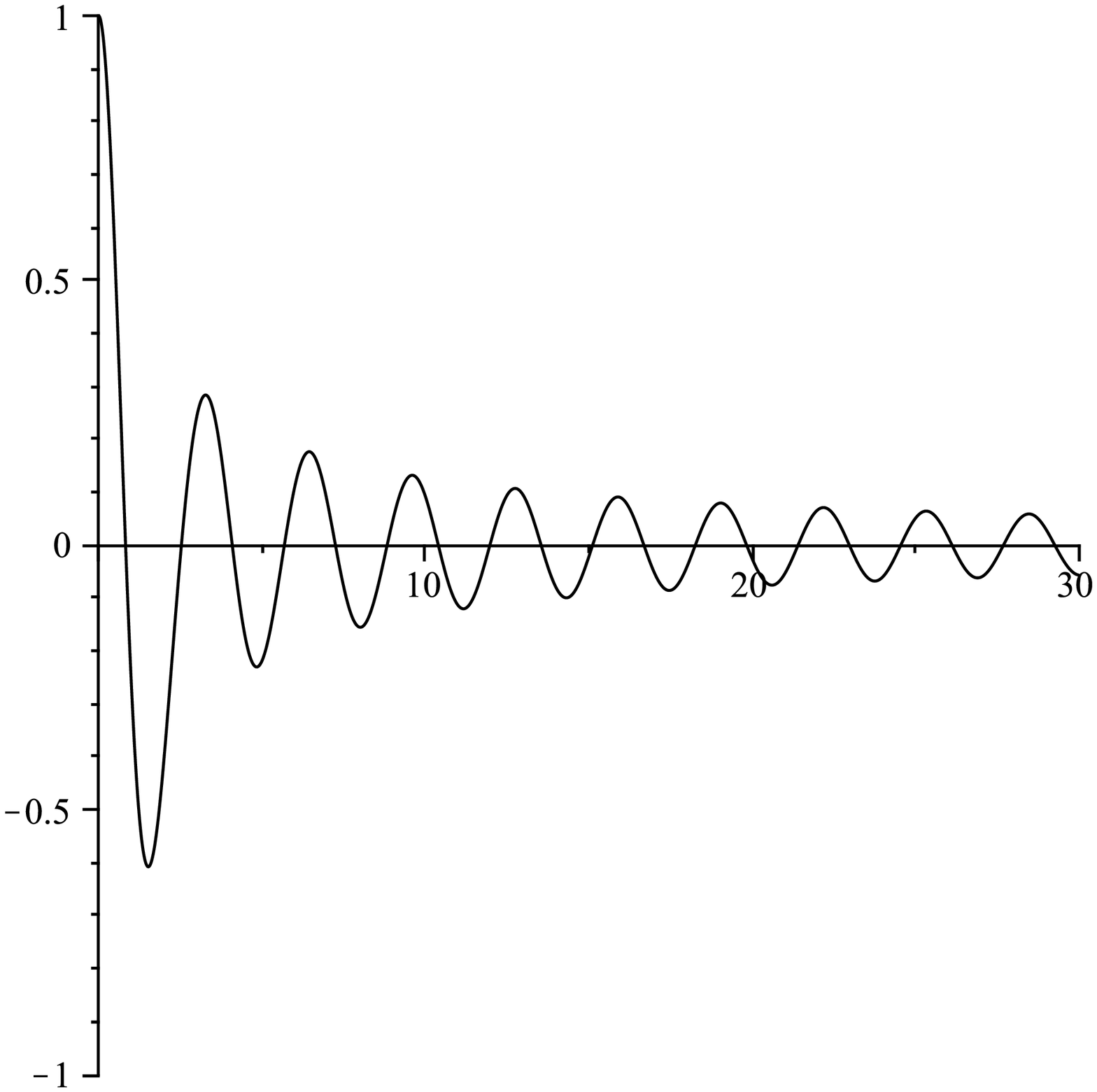}\hfil
\includegraphics[width=0.5\hsize]{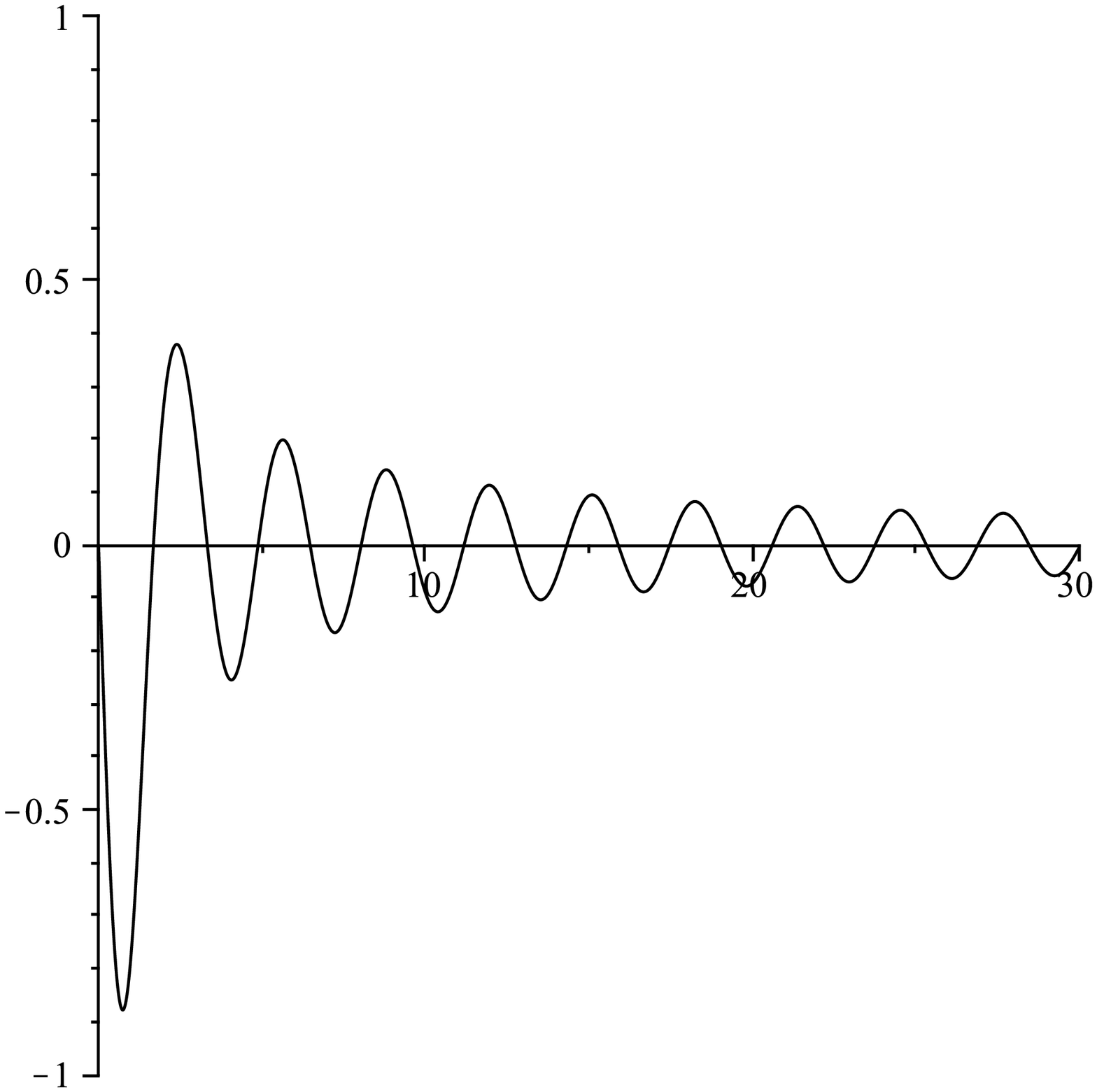}
\end{center}
\caption{$Y_0(t)$ is plotted for $B=J=1$. On the left
$\Re Y_0(t)$ is shown and on the right $\Im X_0(t)$.}
\end{figure}

Finally, we remark that we could have made the plots using only the
Painlev\'e V equation of \cite{MPS1,MPS2,MS1,MS2}. We would, however,
not have been able to calculate the about $10^{11}$ other very accurate
data points
within the ``characteristic triangle" that way.


\subsection{The case of noncritical transverse field $B\ne J$}
\label{sec4b}

In this subsection we assume that $J<B$ in our transverse-field Ising chain,
so that $B^{\ast}<J^{\ast}$ for the dual Ising chain with $B^{\ast}\equiv J$
and $J^{\ast}\equiv B$. We choose time units such that
$B=J^{\ast}=1$,\footnote{Equivalently, we may replace $t$ by
$\bar t\equiv Bt=J^{\ast}t$, $k=J/B$ in the following formulae.} whence
$J=B^{\ast}=k$ and (\ref{dToda}) is to be replaced by
\begin{equation}
\left.\matrix{X_n(t)\ddot X_n(t)-\dot X_n(t)^2&=&
X_{n-1}^{\ast}(t)X_{n+1}^{\ast}(t)-X_n^{\ast}(t)^2\,,\hfill\cr\cr
X_n^{\ast}(t)\ddot X_n^{\ast}(t)-\dot X_n^{\ast}(t)^2&=&
k^2\Big(X_{n-1}(t)X_{n+1}(t)-X_n(t)^2\Big)\,.}\right.
\label{dToda2}
\end{equation}

At the critical point $T=0$, $B=J$, we found it better to work with
the logarithm $\xi_n(t)$ of the pair correlation function $X_n(t)$.
Once we leave the critical point it becomes even essential to do so
because of the exponential decay of the connected pair correlation
with separation $n$, as we need initial conditions for very large $n$.
Therefore, we introduce $\xi_n(t)$ and $\xi_n^{\ast}(t)$ via
\begin{eqnarray}
X_n(t)&=&(1-k^2)^{1/4}\,\mathrm{e}^{\xi_n(t)},\\
X_n^{\ast}(t)&=&(1-k^2)^{1/4}\,(1+\mathrm{e}^{\xi_n^{\ast}(t)}),
\end{eqnarray}
Substituting these into (\ref{dToda2}) these equations become
\begin{eqnarray}
\ddot\xi_n(t)&=&\mathrm{e}^{\xi^{\ast}_{n-1}-2\xi_n}+
\mathrm{e}^{\xi^{\ast}_{n+1}-2\xi_n}-
2\,\mathrm{e}^{\xi^{\ast}_{n}-2\xi_n}\nonumber\\
&&\quad+\,\mathrm{e}^{\xi^{\ast}_{n-1}+\xi^{\ast}_{n+1}-2\xi_n}-
\mathrm{e}^{2\xi^{\ast}_{n}-2\xi_n},
\label{dToda3a}\\\cr
\ddot\xi^{\ast}_n(t)&=&
\frac{k^2(\mathrm{e}^{\xi_{n-1}+\xi_{n+1}-\xi^{\ast}_n}-
\mathrm{e}^{2\xi_{n}-\xi^{\ast}_n})-
{\dot{\bar\xi^{\ast}_n}}^2}{1+\mathrm{e}^{\xi^{\ast}_{n}}}.
\label{dToda3b}
\end{eqnarray}
These expressions were written such that none of the real parts of
their exponents goes rapidly to $+\infty$.

The initial conditions for $\xi_n(t)$ and $\xi_n^{\ast}(t)$ are
determined by (\ref{asyTg}), (\ref{asyTs}), (\ref{init}),
(\ref{initJB}), and (\ref{initBJ}). One immediately sees that
$\xi_n(0)$ and $\xi_n^{\ast}(0)$ roughly grow linearly with $|n|$,
which is much better for the numerics than the exponential behavior
of $X_n(0)$ and $X^{\ast}_n(0)$. Setting, as explained above,
$B=1$ in (\ref{init}) for $\dot X_n(0)$ and $B^{\ast}=k$
for $\dot X^{\ast}_n(0)$, we find
\begin{eqnarray}
\dot\xi_n(0)&=&-\,\frac{2i\,\mathrm{E}(k)}{\pi}\,\delta_{n0},
\label{initxiJB}\\\cr
\dot\xi^{\ast}_n(0)&=&-\,\frac{2i\,\Big(\mathrm{E}(k)-(1-k^2)\,
\mathrm{K}(k)\Big)}{\pi\,\Big(1-(1-k^2)_{\phantom{1}}^{1/4}\Big)}\,\delta_{n0}.
\label{initxiBJ}
\end{eqnarray}
The higher derivatives at $t=0$ can then be evaluated using the pair of
coupled differential equations (\ref{dToda3a}) and (\ref{dToda3b}).

\begin{figure}[tbhp]
\begin{center}
\includegraphics[width=0.5\hsize]{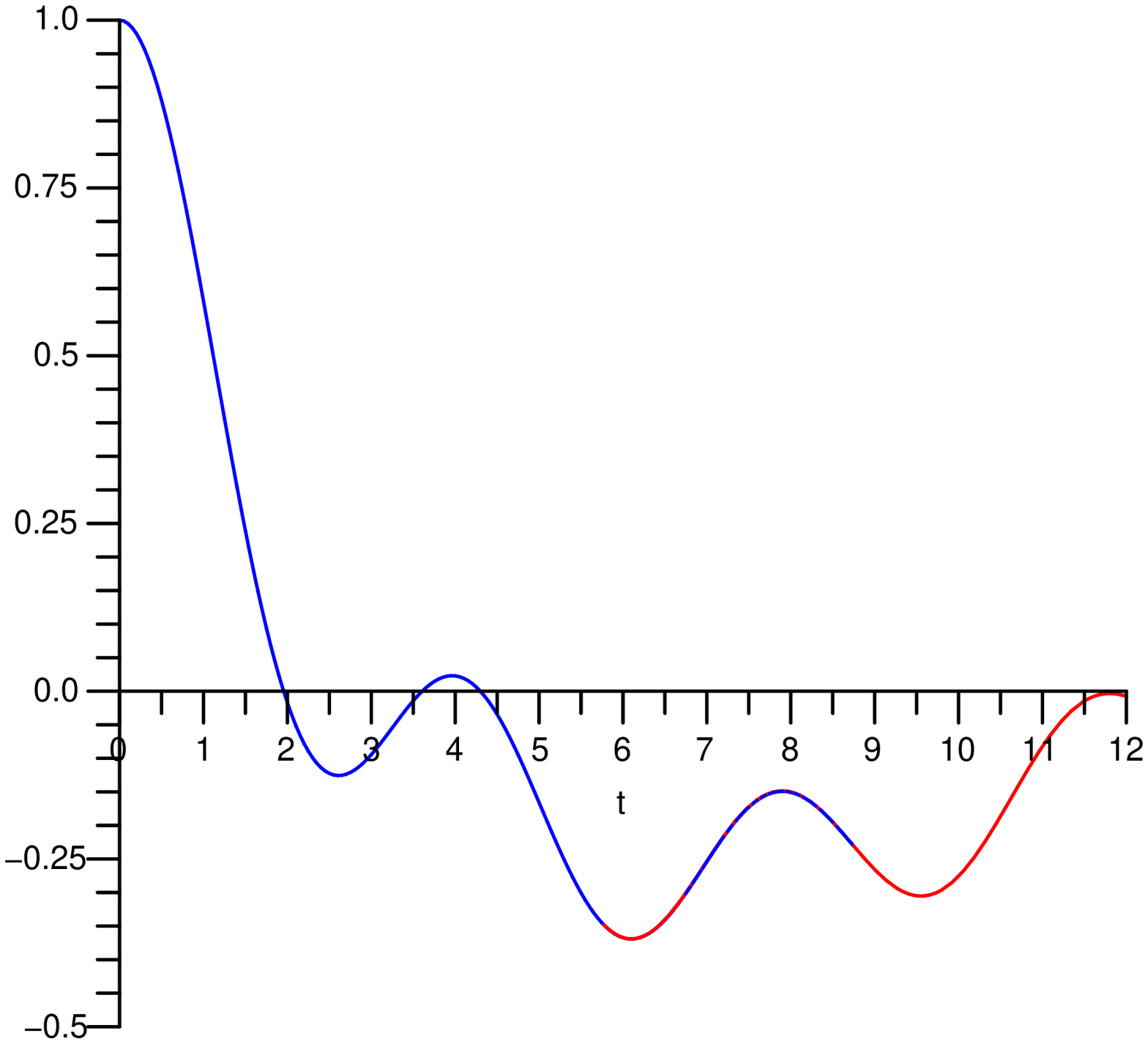}\hfil
\includegraphics[width=0.5\hsize]{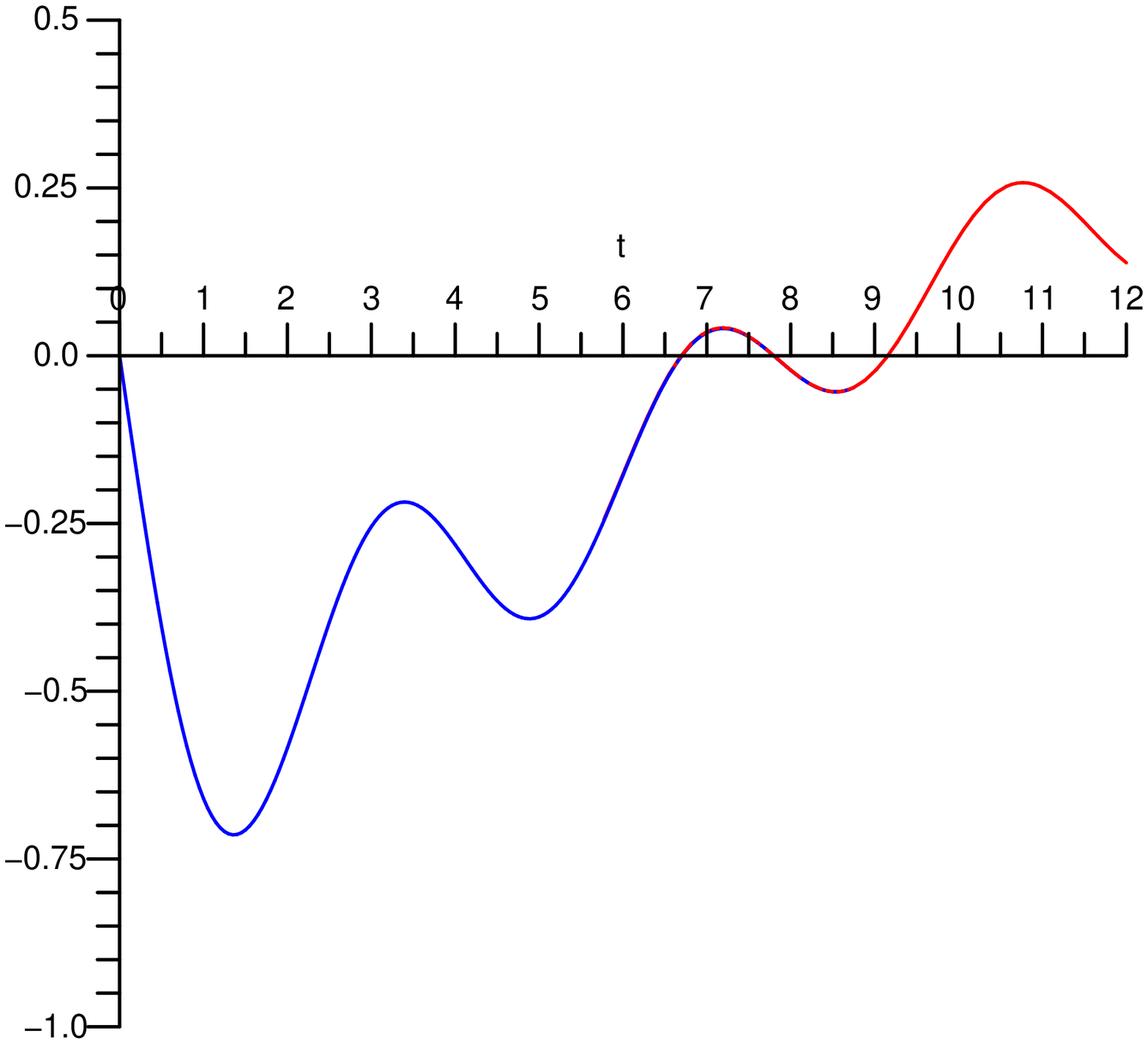}
\end{center}
\caption{$X_0(t)$ for $B=0.7$, $J=1$: $\Re X_0(t)$ and $\Im X_0(t)$.}
\end{figure}
\begin{figure}[tbhp]
\begin{center}
\includegraphics[width=0.5\hsize]{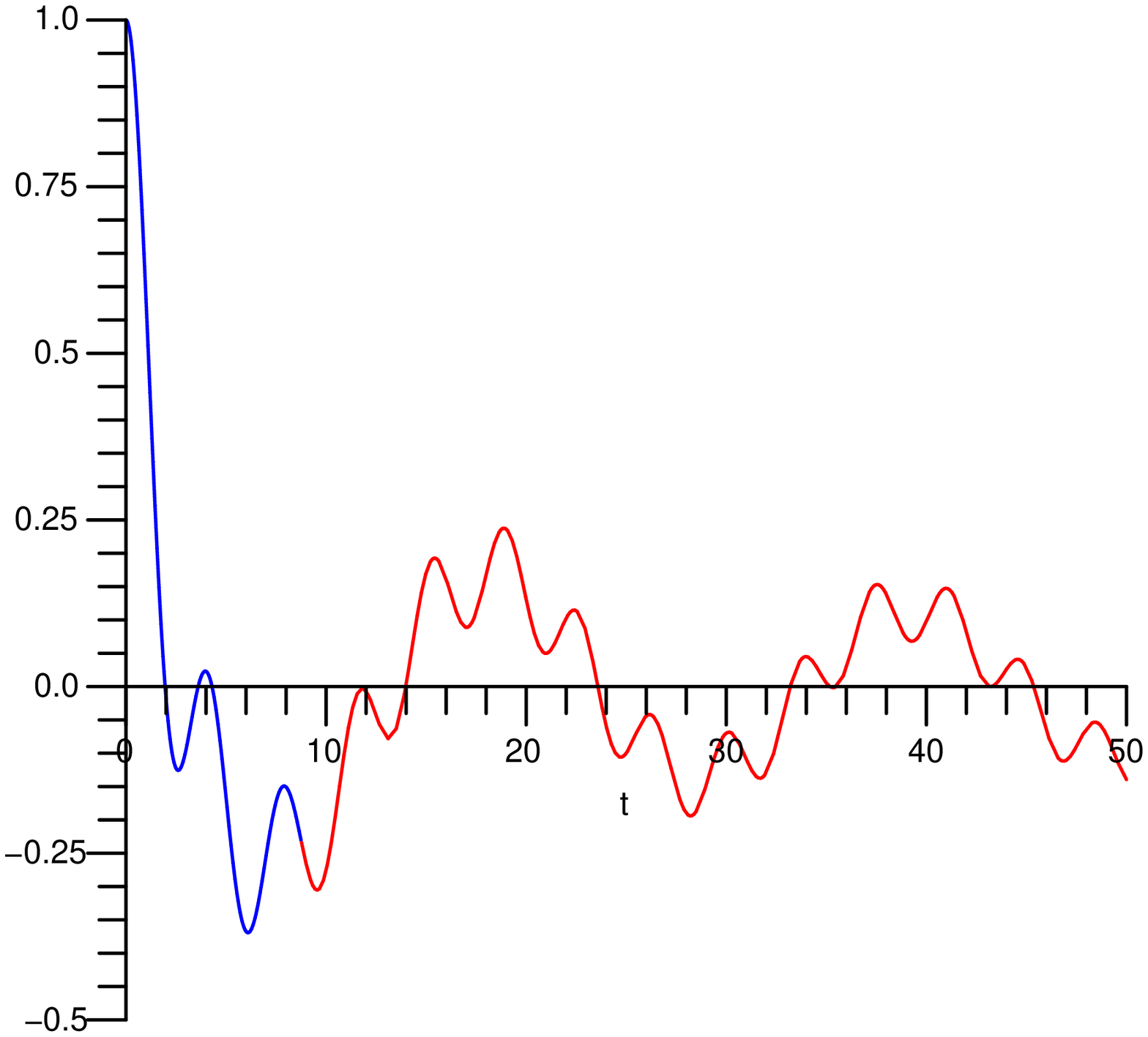}\hfil
\includegraphics[width=0.5\hsize]{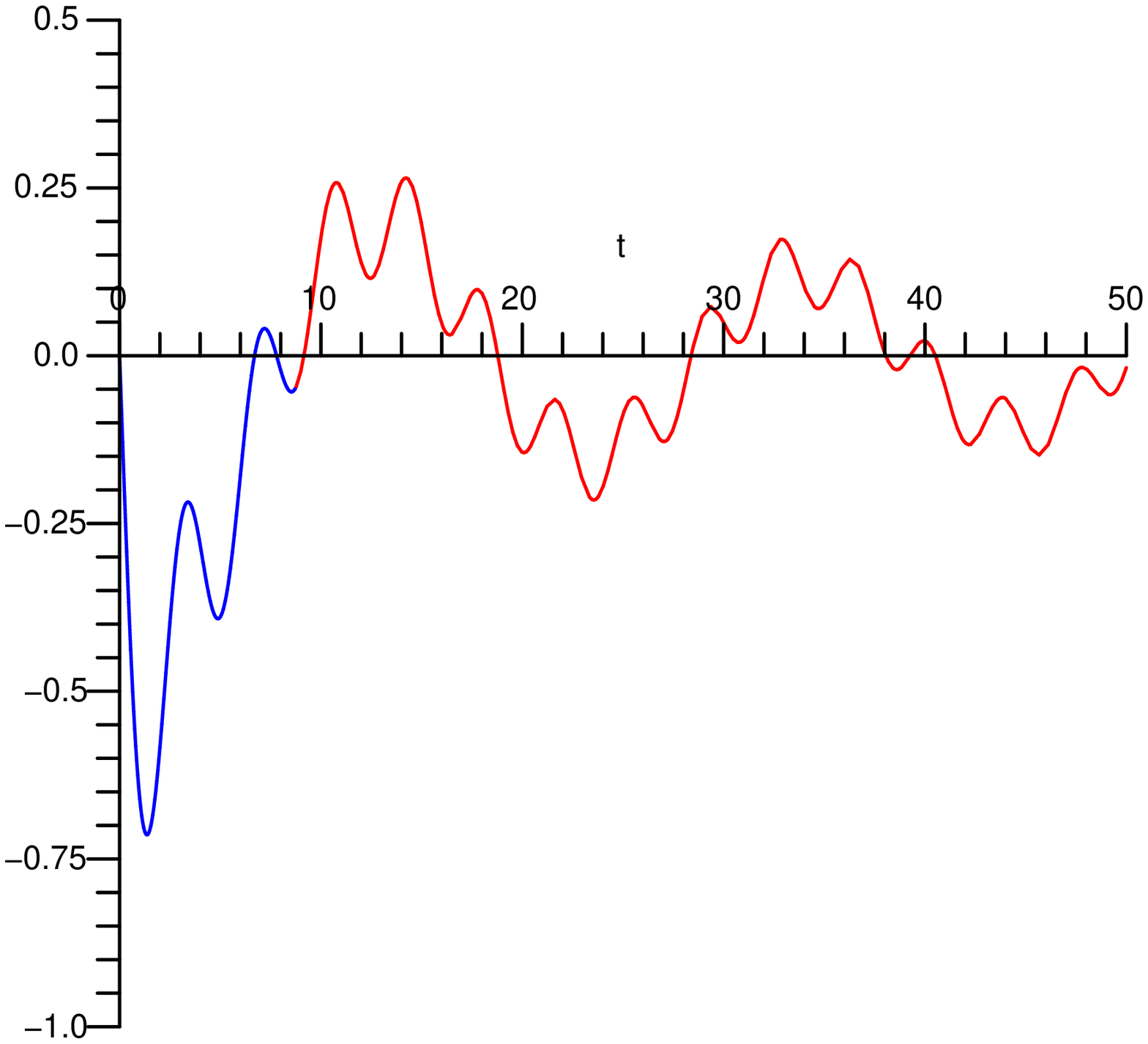}
\end{center}
\caption{$X_0(t)$ for $B=0.7$, $J=1$: $\Re X_0(t)$ and $\Im X_0(t)$.}
\end{figure}
We can now adopt the same strategy as for the critical-field case
$B=J$. Given $\xi_n(t)$ and $\xi_n^{\ast}(t)$ and their first five
time-derivatives at a given $t$, we can calculate
$\xi_n(t+\delta t)$ and $\xi_n^{\ast}(t+\delta t)$ in a Taylor series
through fifth order in $\delta t$ and
$\dot\xi_n(t+\delta t)$ and $\dot\xi_n^{\ast}(t+\delta t)$ in a
Taylor series through fourth order. We have implemented this in a
Maple program with $\delta t=10^{-3}$ going up to $t=8.77$.

Our results for $X_0(t)$ at $k=0.7$ are represented in Fig.~3, using
both our Maple numerical integration results for $t\le8.77$ and our
asymptotic expansions (\ref{asyBgJ}) for $t>5.77$. There is clearly
excellent agreement between the numerical integration and the
asymptotic expansion.\footnote{We note that there is no agreement with
the corresponding lowest-order asymptotic results in \cite{MBA} because
of several misprints there.}

We have also plotted $X_0(t)$ and $Y_0(t)$ for $t\le50$ using the
numerical results for $t\le8.77$ and the asymptotic expansions for
$t\ge8.77$. We have again used (\ref{LazerP}) to calculate $Y_0(t)$.
Note that the lowest frequency, clearly seen in Fig.~4 for $X_0(t)$
is hardly visible after two time derivatives in Fig.~5. The ratio
of the two leading frequencies is 3/17, as follows from (\ref{asyBgJ}),
leading to a suppression factor of 9/289.
\begin{figure}[tbh]
\begin{center}
\includegraphics[width=0.5\hsize]{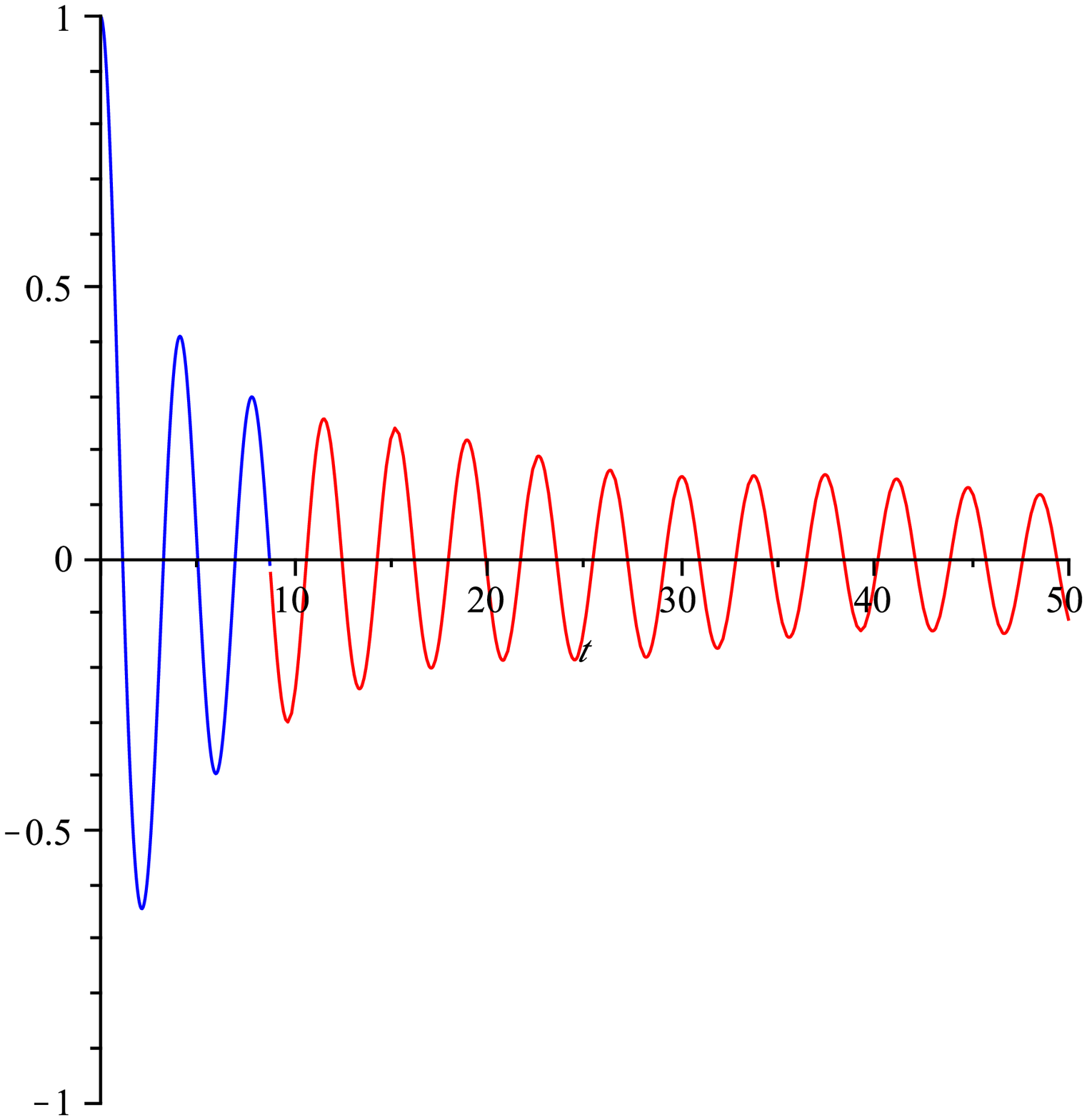}\hfil
\includegraphics[width=0.5\hsize]{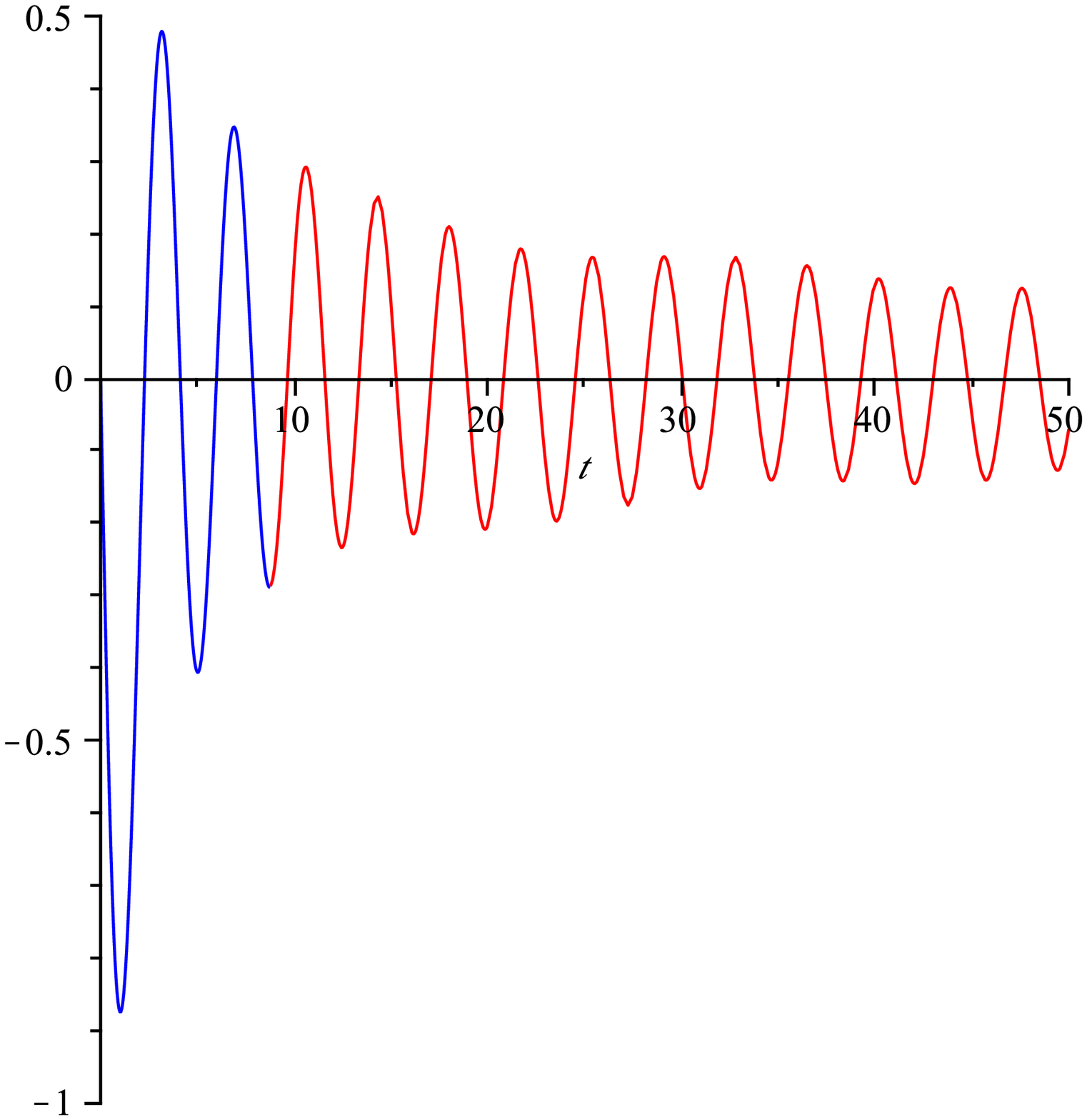}
\end{center}
\caption{$Y_0(t)$ for $B=0.7$, $J=1$: $\Re Y_0(t)$ and $\Im Y_0(t)$.}
\end{figure}

Because of the very nature of the differential equations of \cite{P1}
used, we also got the results for the dual model from the same Maple
run. They correspond to $X_n(t)$ at $B/J=1/k$, or $J/B=0/7$. We have
plotted $X_0(t)$ for $t\le12$ in Fig.~6. Again it is seen that the
numerical results approach the asymptotic behavior well. Note that the
vertical scales are smaller than in Fig.~3, making the small differences
more pronounced.
\begin{figure}[tbhp]
\begin{center}
\includegraphics[width=0.5\hsize]{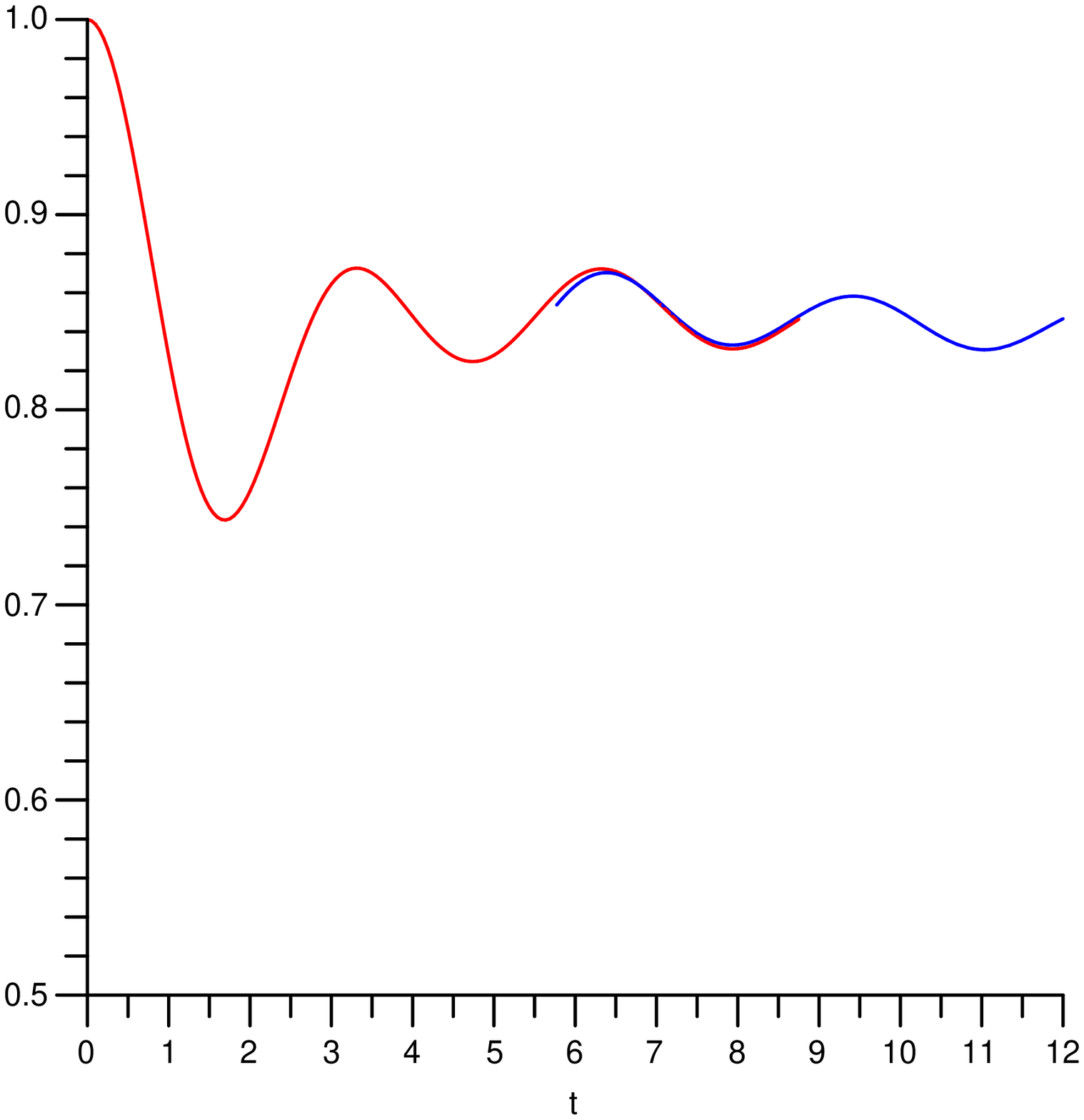}\hfil
\includegraphics[width=0.5\hsize]{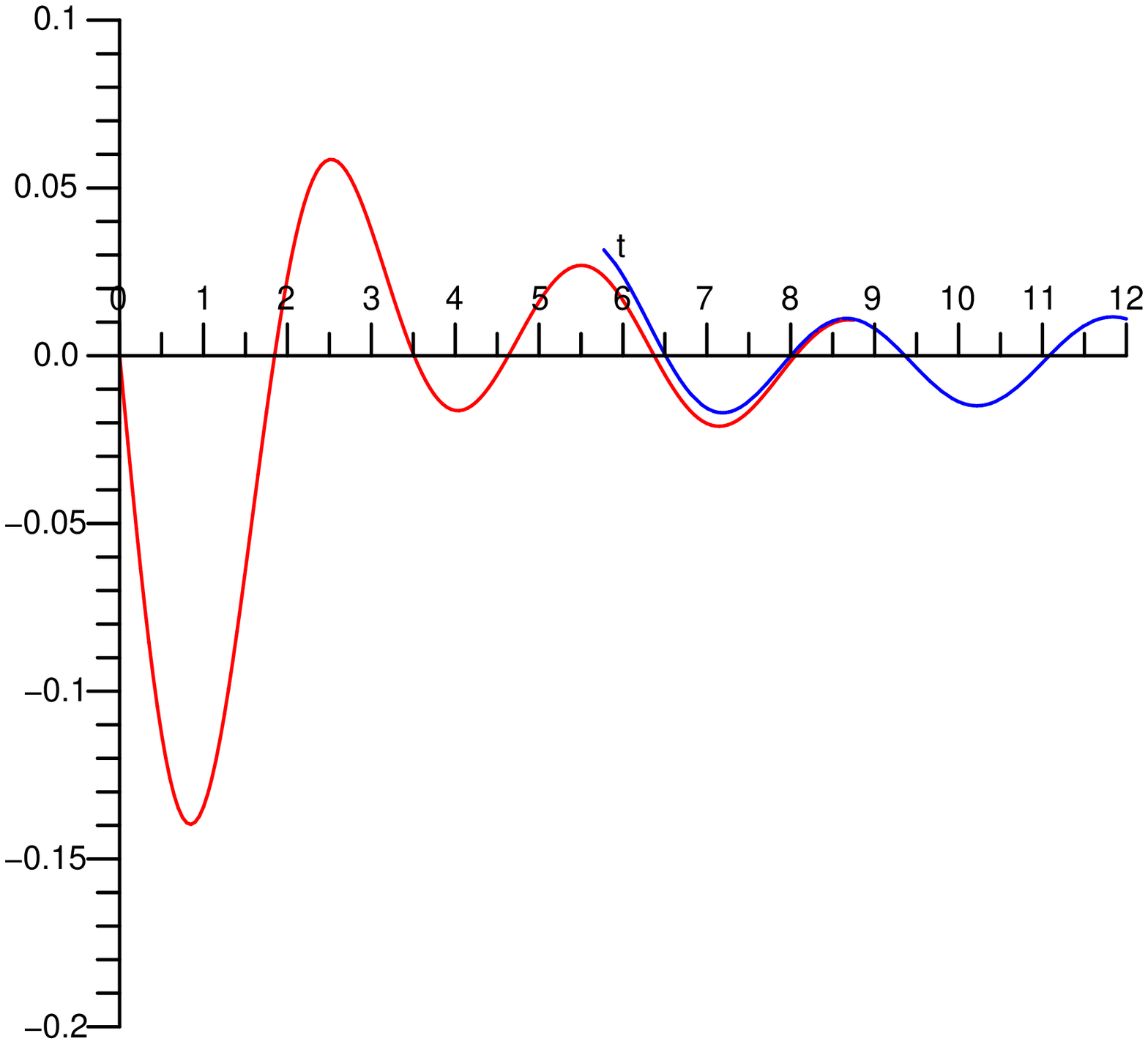}
\end{center}
\caption{$X_0(t)$ for $J=0.7$, $B=1$: $\Re X_0(t)$ and $\Im X_0(t)$.}
\end{figure}
\begin{figure}[tbhp]
\begin{center}
\includegraphics[width=0.5\hsize]{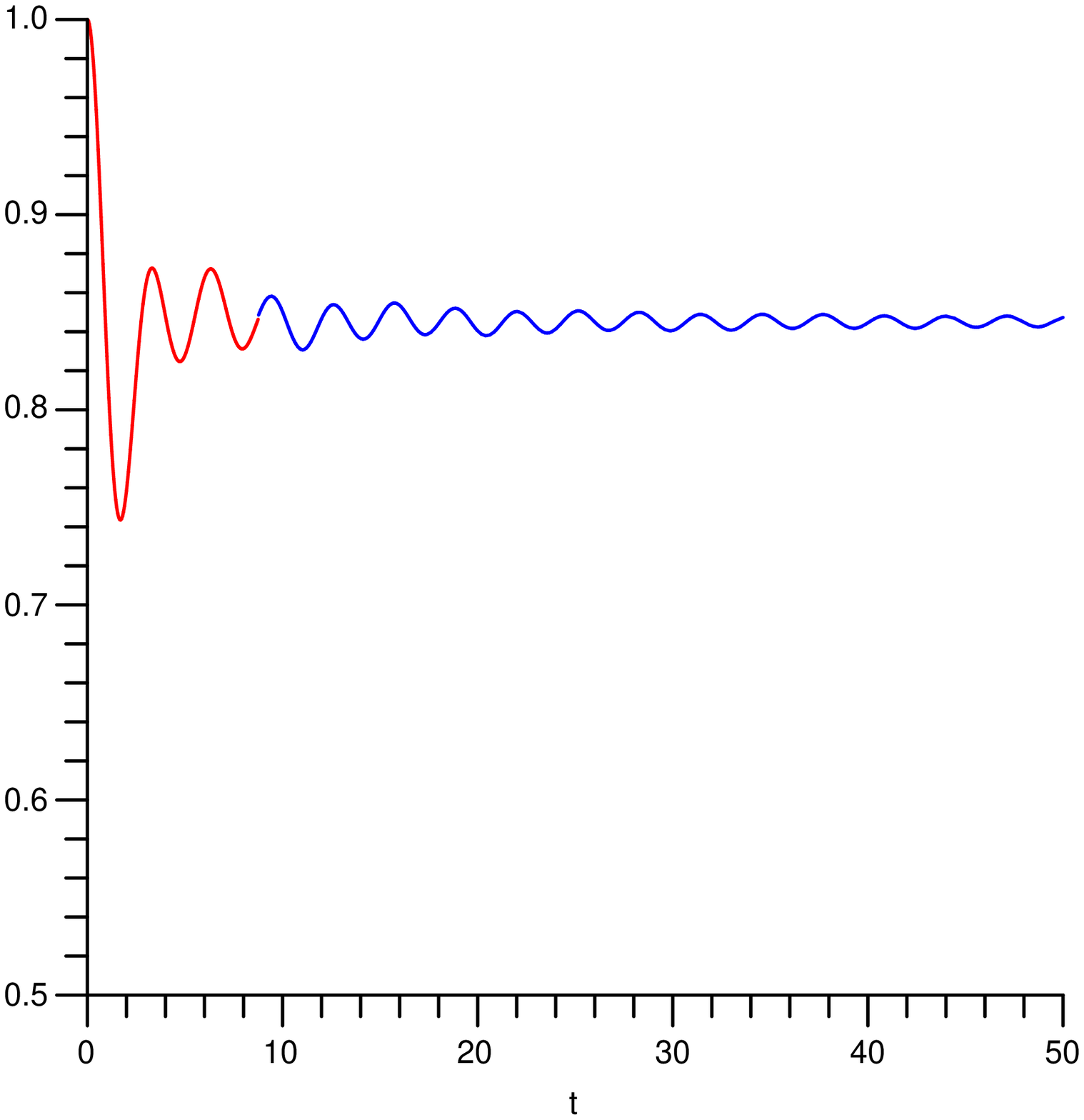}\hfil
\includegraphics[width=0.5\hsize]{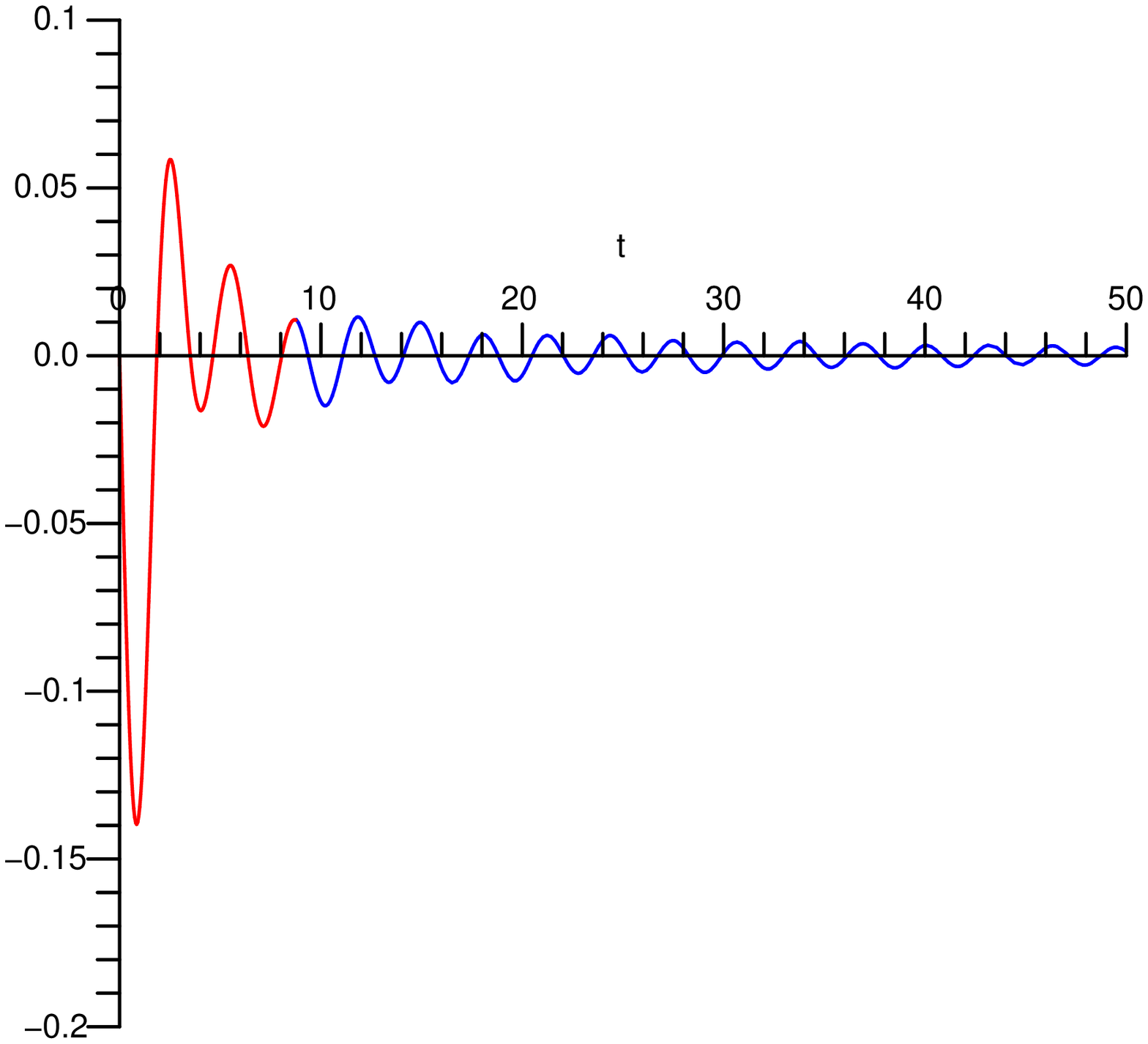}
\end{center}
\caption{$X_0(t)$ for $J=0.7$, $B=1$: $\Re X_0(t)$ and $\Im X_0(t)$.}
\end{figure}
We have also plotted $X_n(t)$ and $Y_n(t)$ for $t\le50$ for the same 
values of the parameters. Unlike Fig.~4, there is now no pronounced
two-frequency behavior, as other frequencies are down two orders in
$t^{-1}$ in (\ref{asyJgB}). In addition, it is clearly seen that $X_0(t)$
decays to the square of the order parameter $(1-k^2)^{1/4}\approx0.845$,
as it should \cite{BM}.

\begin{figure}[tbhp]
\begin{center}
\includegraphics[width=0.5\hsize]{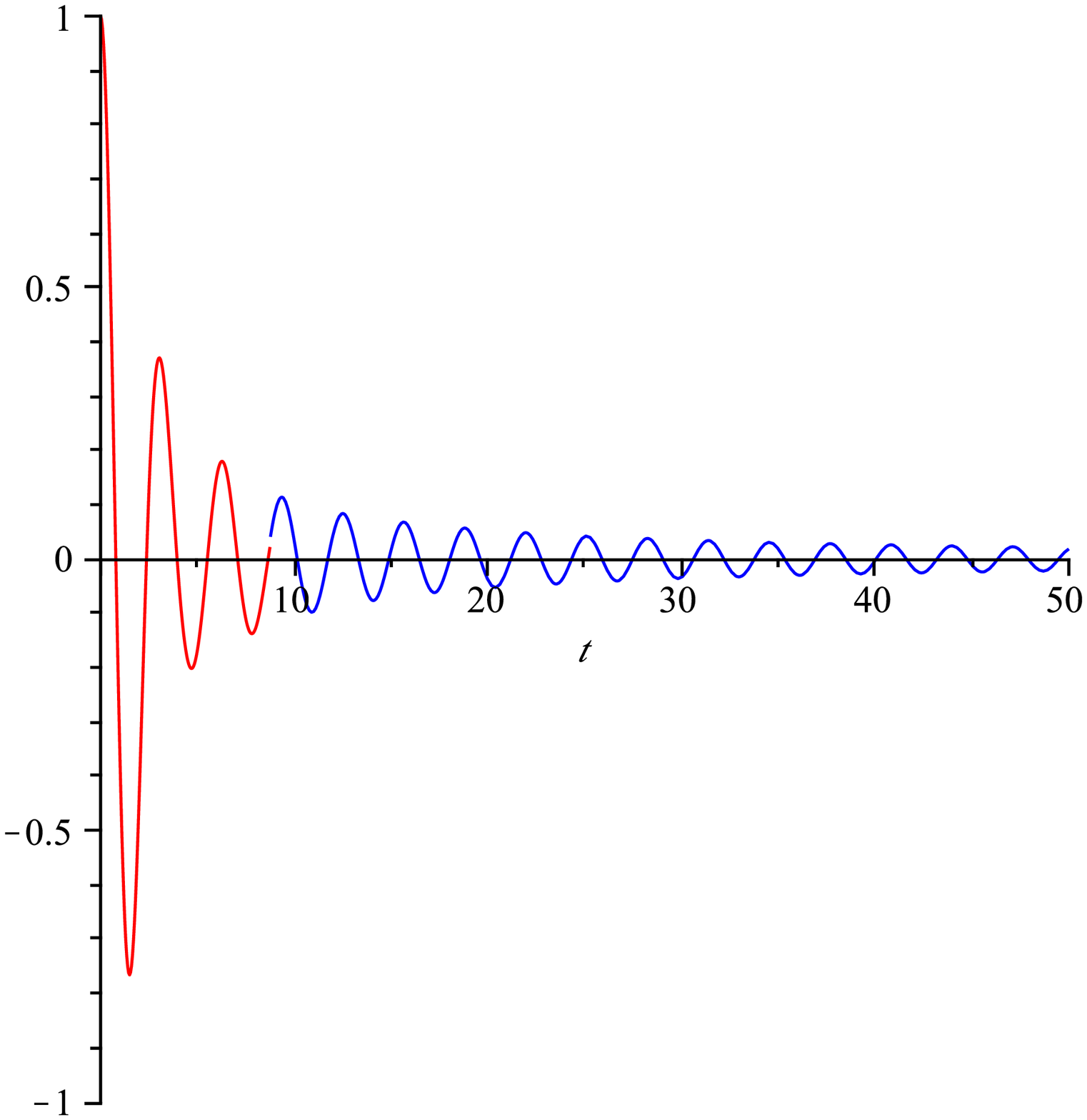}\hfil
\includegraphics[width=0.5\hsize]{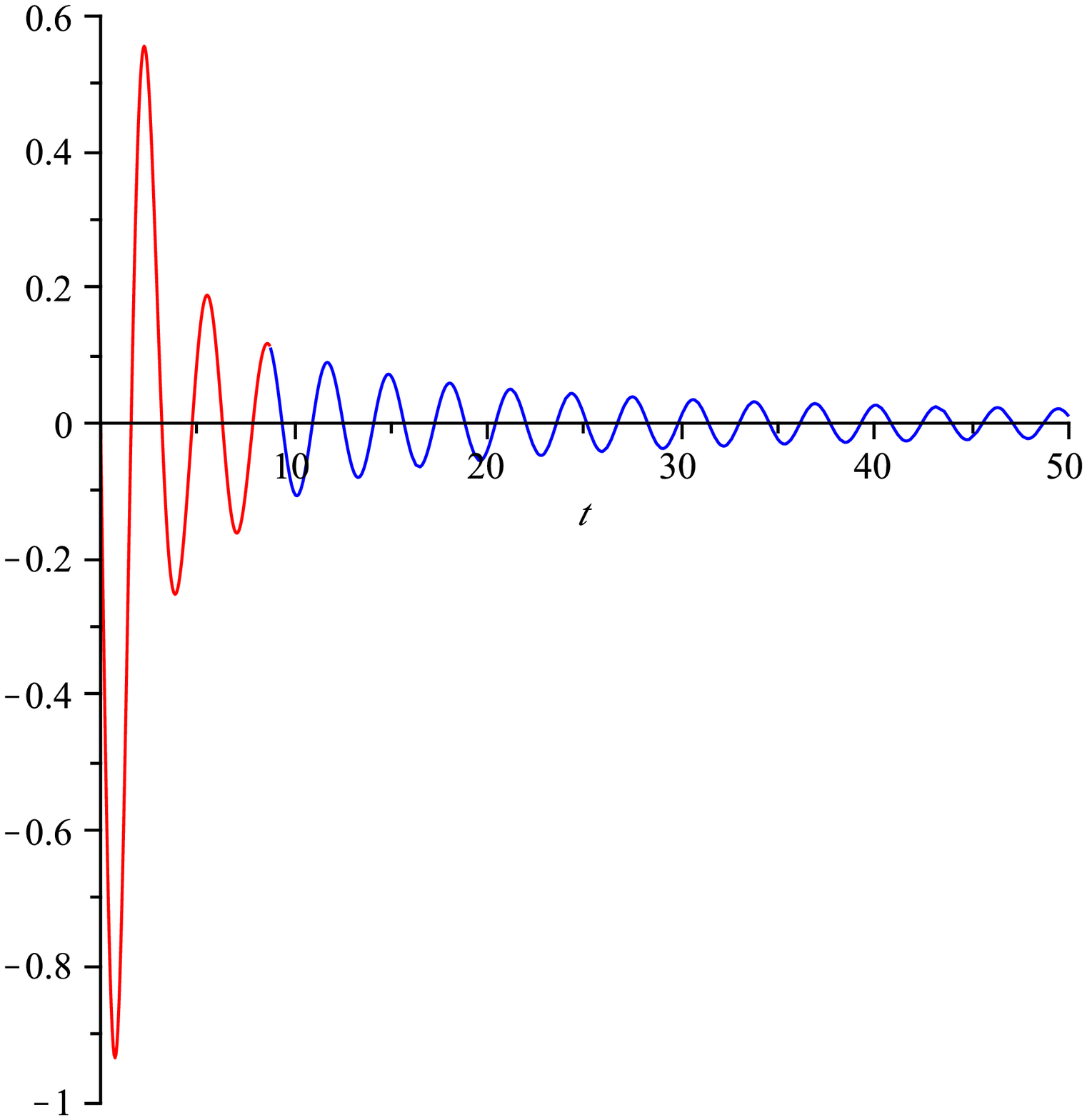}
\end{center}
\caption{$Y_0(t)$ for $J=0.7$, $B=1$: $\Re Y_0(t)$ and $\Im Y_0(t)$.}
\end{figure}

\section{Final Remarks}
\label{sec5}

The original motivation for this work came from two sides. First, we
were told by some colleagues that solving the equations of \cite{P1}
numerically leads to useless results inconsistent with earlier
results, especially \cite{MBA}. We believe that this point of view
has been put to rest by the results in this paper. Secondly, during
a brief visit in 2005 at the University of New South Wales one of us
got to discuss an early version of \cite{HOZ,HOZM} with his colleagues.
It led us to calculate the small and large-$B$ expansions for the
structure function $\langle\sum_j\sigma^x_0\sigma^x_j\rangle$ of
the Ising chain in a transverse field at zero temperature. This turned
out to be nothing but the high- and low-temperature series of the
``diagonal susceptibility" of the two-dimensional Ising model
$\beta\sum_n\langle\sigma_{0\,0}\sigma_{n,n}\rangle$, the calculation
of which is a small part of the calculation of \cite{ONGP1,ONGP2}.
A different approach to the diagonal susceptibility in terms of
form factors and $n$-particle contributions has been given in
\cite{BHMMZ,BBHMWZ} revealing a lot of its mathematical structure.

With the results of this paper it is now also very easy to calculate
extremely accurate values of the $q$-dependence of the structure function
\cite{HOZ,HOZM} and of the second moment correlation length
(as used e.g.\ in \cite{CB}).

Furthermore, it is well-known that the correlation functions of the
zero-field XY-model factor into two correlation functions of the
Ising model in transverse field \cite{PC1,MPS2}, just like
correlations in the two-dimensional square-lattice dimer and
fully-frustrated Ising models factor \cite{AP1,PA}. Therefore, our results
directly apply also to those cases.

Finally, as more and more results on time-correlations in the more
general XXZ model have become available 
\cite{LT,SST,KMST,CM,CHM,PSCHMWA2,HCM,PSCHMWA1},
one may express the hope that some new powerful identities will be
discovered also for this case that will facilitate numerical
calculations dramatically.

\section*{Acknowledgments}

First of all, we must thank the people at the Mathematics and Theoretical
Physics Departments at Australian National University, where this work
was finished, for their hospitality and support. We are also
grateful to Dr.\ C.J.\ Hamer, Dr.\ J.\ Oitmaa and Dr.\ W.\ Zheng
of the University of New South Wales, where the original concept of
this paper was birthed during a brief visit in July, 2005.
Part of this work has been performed on a new computer provided
by Dean Dr.\ P.M.A.\ Sherwood of Arts and Sciences at Oklahoma State
University. This research has also been supported by the
National Science Foundation under Grant No.\ PHY 07-58139 and
by the Australian Research Council under Project ID: LX0989627.




\begin{thebibliography}{10}

\bibitem{Pf}%
Pfeuty, P.:
The one-dimensional Ising model with a transverse field.
Ann. Phys. {\bf 57}, 79--90 (1970).

\bibitem{Na}%
Nambu, Y.:
A note on the eigenvalue problem in crystal statistics.
Progr. Theor. Phys. {\bf 5}, 1--13 (1950).

\bibitem{LSM}%
Lieb, E., Schultz, T., Mattis, D.:
Two soluble models of an antiferromagnetic chain.
Ann. Phys. {\bf 16}, 407--466 (1961).

\bibitem{Ka}%
Katsura, S.:
Statistical mechanics of the anisotropic linear Heisenberg model.
Phys. Rev. {\bf 127}, 1508--1518, 2835 (1962).

\bibitem{Ni}%
Niemeijer, Th.:
Some exact calculations on a chain of spins $\frac12$.
Physica {\bf 36}, 377--419 (1967).

\bibitem{KHS}%
Katsura, S., Horiguchi, T., Suzuki, M.:
Dynamical properties of the isotropic $XY$ model.
Physica {\bf 46}, 67--86 (1970).

\bibitem{TH}%
Tommet, T.N., Huber, D.L.:
Dynamical correlation functions of the transverse spin and
energy density for the one-dimensional spin-1/2 Ising model
with a transverse field.
Phys. Rev. B {\bf 11}, 450--457 (1975).

\bibitem{HT}%
Huber, D.L., Tommet, T.:
Spin and energy coupling in the Ising model with a transverse field:
One dimension, $T=\infty$.
Solid State Commun. {\bf 13}, 1973--1976 (1973).

\bibitem{PCS}%
Perk, J.H.H., Capel, H.W., Siskens, Th.J.:
Time-correlation functions and ergodic properties in the
alternating XY-chain.
Physica A {\bf 89}, 304--325 (1977).

\bibitem{PM}%
Pesch, W., Mikeska, H.J.:
Dynamical correlation functions in the $x$-$y$ model.
Z. Phys. B {\bf 30}, 177--182 (1978).

\bibitem{MBA}%
McCoy, B.M., Barouch, E., Abraham, D.B.:
Statistical mechanics of the $XY$ model. IV.
Time-dependent spin-correlation functions.
Phys. Rev. A {\bf 4}, 2331--2341 (1971).

\bibitem{P1}%
Perk, J.H.H.:
Equations of motion for the transverse correlations of the one-dimensional
XY-model at finite temperature.
Phys. Lett. A {\bf 79}, 1--2 (1980).

\bibitem{PCQN}%
Perk, J.H.H., Capel, H.W., Quispel, G.R.W., Nijhoff, F.W.:
Finite-temperature correlations for the Ising chain in a transverse field.
Physica A {\bf 123}, 1--49 (1984).

\bibitem{MW2}%
McCoy, B.M., Wu, T.T.:
Nonlinear partial difference equations for the
two-dimensional Ising model.
Phys. Rev. Lett. {\bf 45}, 675--678 (1980).

\bibitem{P2}%
Perk, J.H.H.:
Quadratic identities for Ising model correlations.
Phys. Lett. A {\bf 79}, 3--5 (1980).

\bibitem{SJL}%
Sur, A., Jasnow, D., Lowe, I.J.:
Spin dynamics for the one-dimen\-sional $XY$ model at
infinite temperature.
Phys. Rev. B {\bf 12}, 3845--3848 (1975).

\bibitem{BJ1}%
Brandt, U., Jacoby, K.:
Exact results for the dynamics of one-dimen\-sional spin-systems.
Z. Phys. B {\bf 25}, 181--187 (1976).

\bibitem{CP}%
Capel, H.W., Perk, J.H.H.:
Autocorrelation function of the x-compon\-ent of the magnetization in the
one-dimen\-sional XY-model.
Physica A {\bf 87}, 211--242 (1977).

\bibitem{BJ2}%
Brandt, U., Jacoby, K.:
The transverse correlation function of aniso\-tropic $X$$-$$Y$-chains:
Exact results at $T=\infty$.
Z. Phys. B {\bf 26}, 245--252 (1977).

\bibitem{PC1}%
Perk, J.H.H., Capel, H.W.:
Time-dependent xx-correlations in the one-dimen\-sional XY-model.
Physica A {\bf 89}, 265--303 (1977).

\bibitem{PC2}%
Perk, J.H.H., Capel, H.W.:
Transverse correlations in the inhomogeneous XY-model at infinite temperature.
Physica A {\bf 92}, 163--184 (1978).

\bibitem{PC3}%
Perk, J.H.H., Capel, H.W.:
Time- and frequency-dependent correlation functions for the homogeneous
and alternating XY-models.
Physica A {\bf 100}, 1--23 (1980).

\bibitem{SVM}%
Stolze, J., Viswanath, V.S., M\"uller, G.:
Dynamics of semi-infinite quantum spin chains at $T=\infty$.
Z. Phys. B {\bf 89}, 45--55 (1992).

\bibitem{JM}%
Johnson, J.D., McCoy, B.M.:
Off-diagonal time-dependent spin-corre\-lation functions
of the $XY$ model.
Phys. Rev. A {\bf 4}, 2314--2324 (1971).

\bibitem{VT}%
Vaidya, H.G., Tracy, C.A.:
Transverse time-dependent spin-corre\-lation functions
for the one-dimensional $XY$ model at zero temperature.
Physica A {\bf 92}, 1--41 (1978).

\bibitem{MPS1}%
McCoy, B.M., Perk, J.H.H., Shrock, R.E.:
Time-dependent correlation functions of the transverse Ising chain
at the critical magnetic field.
Nucl. Phys. B {\bf 220} [FS8], 35--47 (1983).

\bibitem{MPS2}%
McCoy, B.M., Perk, J.H.H., Shrock, R.E.:
Correlation functions of the transverse Ising chain at
the critical field for large temporal and spatial separations.
Nucl. Phys. B {\bf 220} [FS8], 269--282 (1983).

\bibitem{MS1}%
M\"uller, G., Shrock, R.E.:
Dynamic correlation functions for quantum spin chains.
Phys. Rev. Lett. {\bf 51}, 219--222 (1983).

\bibitem{MS2}%
M\"uller, G., Shrock, R.E.:
Dynamic correlation functions for one-dimensional
quantum-spin systems: New results based on a rigorous
approach.
Phys. Rev. B {\bf 29}, 288--301 (1984).

\bibitem{MS3}%
M\"uller, G., Shrock, R.E.:
Susceptibilities of one-dimensional
quantum spin models at zero temperature.
Phys. Rev. B {\bf 30}, 5254--5264 (1984).

\bibitem{MS4}%
M\"uller, G., Shrock, R.E.:
Wave-number-dependent susceptibilities of one-dimensional
quantum spin models at zero temperature.
Phys. Rev. B {\bf 31}, 637--640 (1985).

\bibitem{IIKS1}%
Its, A.R., Izergin, A.G., Korepin, V.E., Slavnov, N.A.:
Differential equations for quantum correlation functions.
Int. J. Mod. Phys. B {\bf 4}, 1003--1037 (1990).

\bibitem{IIKN}%
Its, A.R., Izergin, A.G., Korepin, V.E., Novokshenov, V.Ju.:
Temperature autocorrelations of the transverse Ising chain
at the critical magnetic field.
Nucl. Phys. B {\bf 340}, 752--758 (1990).

\bibitem{CIKT}%
Colomo, F., Izergin, A.G., Korepin, V.E., Tognetti, V.:
Temperature correlation functions in the XX0 Heisenberg chain. I.
Teor. Mat. Fiz. {\bf 94}, 19--51 (1993)
[Theor. Math. Phys. {\bf 94}, 11--38 (1993)].

\bibitem{IIKS2}%
Its, A.R., Izergin, A.G., Korepin, V.E., Slavnov, N.A.:
Temperature correlations of quantum spins.
Phys. Rev. Lett. {\bf 70}, 1704--1706, 2357 (1993).

\bibitem{IIKS3}%
Its, A.R., Izergin, A.G., Korepin, V.E., Slavnov, N.A.:
Integrable differential equations for temperature correlation
functions of the XXO Heisenberg chain.
Zap. Nauch. Sem. POMI {\bf 205}, 6--20 (1993)
[J. Math. Sciences {\bf 80}, 1747--1759 (1996)].

\bibitem{DZ}%
Deift, P., Zhou, X.:
Long-time asymptotics for the autocorrelation function of
the transverse Ising chain at the critical magnetic field.
In: Singular Limits of Dispersive Waves (Lyon, 1991),
NATO Adv. Sci. Inst. Ser. B Phys., Vol. 320, pp. 183--201
Plenum, New York (1994).

\bibitem{SNM}%
Stolze, J., N\"oppert, A., M\"uller, G.:
Gaussian, exponential, and power-law decay of
time-dependent correlation functions in quantum spin chains.
Phys. Rev. B {\bf 52}, 4319--4326 (1995).
arXiv:cond-mat/9501079.

\bibitem{Sa1}%
Sachdev, S.:
Universal, finite temperature, crossover functions of the quantum
transition in the Ising chain in a transverse field.
Nucl. Phys. B {\bf 464}, 576--595 (1996).

\bibitem{Sa2}%
Sachdev, S.:
Finite temperature correlations in the
one-dimensional quantum Ising model.
Nucl. Phys. B {\bf 482}, 579--612 (1996).

\bibitem{DG}%
Doyon, B., Gamsa, A.:
Integral equations and long-time asymptotics for finite-temperature
Ising chain correlation functions.
J. Stat. Mech. P03012, 40 pp. (2008).
arXiv:0711.4619.

\bibitem{JM17}%
Jimbo, M., Miwa, T.:
Studies on holonomic quantum fields. XVII.
Proc. Japan Acad. A {\bf 56}, 405--410 (1980).
Errata {\bf 57}, 347 (1987).

\bibitem{Wi}%
Witte, N.S.:
Isomonodromic deformation theory and the
next-to-diagonal correlations of the anisotropic square
lattice Ising model.
J. Phys. A {\bf 40}, F491--F501 (2007).

\bibitem{Wu}%
Wu, T.T.:
Theory of Toeplitz determinants and the spin correlations
of the two-dimensional Ising model. I.
Phys. Rev. {\bf 149}, 380--401 (1966).

\bibitem{MW1}%
McCoy, B.M., Wu, T.T.:
The Two-Dimensional Ising Model.
Harvard University Press, Cambridge, Massachusetts (1973).

\bibitem{AP4}%
Au-Yang, H., Perk, J.H.H.:
Correlation functions and susceptibility in the $Z$-invariant Ising model.
In: Kashiwara, M., Miwa, T. (eds.) MathPhys Odyssey 2001:
Integrable Models and Beyond, pp. 23--48, Birkh\"auser, Boston (2002).

\bibitem{LM}%
Li, N.Y., Mansour, T.:
An identity involving Narayana numbers.
European J. Combin. {\bf 29}, 672--675 (2008).

\bibitem{Gh}%
Ghosh, R.K.:
On the low-temperature series expansion for the diagonal
correlation functions in the two-dimensional Ising model.
arXiv:cond-mat/0505166 (7 pp.).

\bibitem{ONGP1}%
Orrick, W.P., Nickel, B., Guttmann, A.J., Perk, J.H.H.:
The susceptibility of the square lattice Ising model: New developments.
J. Stat. Phys. {\bf 102}, 795--841 (2001). arXiv:cond-mat/0103074.
See http://www.ms.unimelb.edu.au/\~{}tonyg
for the complete set of series coefficients.


\bibitem{ONGP2}%
Orrick, W.P., Nickel, B.G., Guttmann, A.J., Perk, J.H.H.:
Critical behavior of the two-dimensional Ising susceptibility.
Phys. Rev. Lett. {\bf 86}, 4120--4123 (2001).
arXiv:cond-mat/0009059.

\bibitem{FB}%
Fisher, M.E., Burford, R.J.:
Theory of critical-point scattering and correlations.
I.~The Ising model.
Phys. Rev. {\bf 156}, 583--622 (1967). See footnote 25
on p.~591.

\bibitem{WMTB}%
Wu, T.T., McCoy, B.M., Tracy, C.A., Barouch, E.:
Spin-spin correlation functions for the
two-dimensional Ising model:
Exact theory in the scaling region.
Phys. Rev. B {\bf 13}, 316--374 (1976).

\bibitem{KAP1}%
Kong, X.-P., Au-Yang, H., Perk, J.H.H.:
 New results for the susceptibility of the two-dimensional Ising model
at criticality.
Phys. Lett. A {\bf 116}, 54--56 (1986).

\bibitem{KAP2}%
Kong, X.-P., Au-Yang, H., Perk, J.H.H.:
Logarithmic singularities of $Q$-dependent susceptibility
of 2-d Ising model.
Phys. Lett. A {\bf 118}, 336--340 (1986).

\bibitem{KAP3}%
Kong, X.-P., Au-Yang, H., Perk, J.H.H.:
Comment on a paper by Yamada and Suzuki.
Progr. Theor. Phys. {\bf 77}, 514--516 (1987).

\bibitem{Ko}%
Kong, X.-P.:
Wave-Vector Dependent Susceptibility of the
Two-Dimen\-sio\-nal Ising Model.
Ph.D. Thesis, State University of New York at Stony Brook
(September, 1987).

\bibitem{AJP}%
Au-Yang, H., Jin, B.-Q., Perk, J.H.H.:
Wavevector-dependent susceptibility in quasiperiodic Ising models.
J. Stat. Phys. {\bf 102}, 501--543 (2001).

\bibitem{AP3}%
Au-Yang, H., Perk, J.H.H.:
Wavevector-dependent susceptibility in aperiodic planar Ising models.
In: Kashiwara, M., Miwa, T. (eds.) MathPhys Odyssey 2001:
Integrable Models and Beyond, pp. 1--21, Birkh\"auser, Boston (2002).

\bibitem{AP7}%
Au-Yang, H., Perk, J.H.H.:
$Q$-dependent susceptibilities in $Z$-invariant pentagrid Ising models.
J. Stat. Phys. {\bf 127}, 221--264 (2007).
arXiv:cond-mat/0409557.

\bibitem{AP8}%
Au-Yang, H., Perk, J.H.H.:
$Q$-dependent susceptibilities in ferromagnetic
quasiperiodic $Z$-invariant Ising models.
J. Stat. Phys. {\bf 127}, 265--286 (2007).
arXiv:cond-mat/0606301.

\bibitem{Bax}%
Baxter, R.J.:
Solvable eight-vertex model on an arbitrary planar lattice,
Philos. Trans. Roy. Soc. London Ser. A {\bf 289}, 315--346 (1978).

\bibitem{AP2}%
Au-Yang, H., Perk, J.H.H.:
Critical correlations in a $Z$-invariant inhomogeneous Ising model.
Physica A {\bf 144}, 44--104 (1987).

\bibitem{AP5}%
Au-Yang, H., Perk, J.H.H.:
New results for susceptibilities in planar Ising models.
Int. J. Mod. Phys. B {\bf 16}, 2089--2095 (2002).

\bibitem{AP6}%
Au-Yang, H., Perk, J.H.H.:
Susceptibility calculations in periodic and
quasiperiodic planar Ising models.
Physica A {\bf 321}, 81--89 (2003).

\bibitem{MT}%
McCoy, B.M., Tang, S.:
Connection formulae for Painlev\'e V functions.
Physica D {\bf 19}, 42--72 (1986).

\bibitem{BM}%
Barouch, E., McCoy, B.M.:
Statistical mechanics of the $XY$ model. II.
Spin-correlation functions.
Phys. Rev. A {\bf 3}, 786--804 (1971).

\bibitem{LP}%
Lajzerowicz, J., Pfeuty, P.:
Space-time--dependent spin correlation of the 
one-dimensional Ising model with a transverse field.
Application to higher dimension.
Phys. Rev. B {\bf 11}, 4560--4562 (1975).

\bibitem{HOZ}%
Hamer, C.J., Oitmaa, J., Zheng, W.:
One-particle dispersion and spectral weights in the
transverse Ising model.
Phy. Rev. B {\bf 74}, 174428, 10 pp. (2006).

\bibitem{HOZM}%
Hamer, C.J., Oitmaa, J., Zheng, W., McKenzie, R.H.:
Critical behavior of one-particle spectral weights
in the transverse Ising model.
Phys. Rev. B {\bf 74}, 060402(R), 4 pp. (2006).

\bibitem{BHMMZ}%
Boukraa, S., Hassani, S., Maillard, J.-M., McCoy, B.M., Zenine, N.:
The diagonal Ising susceptibility.
J. Phys. A: Math. Theor. {\bf 40}, 8219--8236 (2007).
arXiv:math-ph/0703009.

\bibitem{BBHMWZ}%
Bostan, A., Boukraa, S., Hassani, S., Maillard, J.-M., Weil, J.-A.,
Zenine, N.:
Globally nilpotent differential operators and the
square Ising model.
J. Phys. A: Math. Theor. {\bf 42}, 125206, 50 pp. (2009).
arXiv:0812.4931.

\bibitem{CB}%
Campbell, I.A., Butera, P.:
Extended scaling for the high-dimension and square-lattice
Ising ferromagnets.
Phys. Rev. B {\bf 78}, 024435, 7 pp. (2008).

\bibitem{AP1}%
Au-Yang, H., Perk, J.H.H.:
Ising correlations at the critical temperature.
Phys. Lett. A {\bf 104}, 131--134 (1984).

\bibitem{PA}%
Perk, J.H.H., Au-Yang, H.:
Some recent results on pair correlation functions
and susceptibilities in exactly solvable models.
J. Phys.: Conf. Ser. {\bf 42}, 231--238 (2006).
arXiv:math-ph/0606046.

\bibitem{LT}%
Lukyanov, S., Terras, V.:
Long-distance asymptotics of spin-spin correlation
functions for the XXZ spin chain.
Nucl. Phys. B {\bf 654} [FS], 323--356 (2003).
arXiv:hep-th/0206093.

\bibitem{SST}%
Sato, J., Shiroishi, M., Takahashi, M.:
Evaluation of dynamic spin structure factor
for the spin-1/2 XXZ chain in a magnetic field.
J. Phys. Soc. Japan {\bf 73}, 3008--3014 (2004).
arXiv:cond-mat/0410102.

\bibitem{KMST}%
Kitanine, N., Maillet, J.M., Slavnov, N.A., Terras, V.:
Dynamical correlation functions of the $XXZ$ spin-1/2 chain.
Nucl. Phys. B {\bf 729} [FS], 558--580 (2005).
arXiv:hep-th/0407108.

\bibitem{CM}%
Caux, J.-S., Maillet, J.-M.:
Computation of dynamical correlation\break functions of Heisenberg
chains in a magnetic field.
Phys. Rev. Lett. {\bf 95}, 077201, 3 pp. (2005).
arXiv:cond-mat/0502365.

\bibitem{CHM}%
Caux, J.-S., Hagemans, R., Maillet, J.-M.:
Computation of dynamical correlation functions of Heisenberg chains:
the gapless anisotropic regime,
J. Stat. Mech. P09003, 20 pp. (2005).
arXiv:cond-mat/0506698.

\bibitem{PSCHMWA2}%
Pereira, R.G., Sirker, J., Caux, J.-S., Hagemans, R.,
Maillet, J.M., White, S.R., Affleck, I.:
The dynamical spin structure factor for the anisotropic
spin-1/2 Heisenberg chain.
Phys. Rev. Lett. {\bf 96}, 257202, 4 pp. (2006).
arXiv:cond-mat/0603681.

\bibitem{HCM}%
Hagemans, R., Caux, J.-S., Maillet, J.M.:
How to calculate correlation functions of
Heisenberg chains.
AIP Conf. Proc. {\bf 846}, 245--254 (2006).
arXiv:cond-mat/0611467.

\bibitem{PSCHMWA1}%
Pereira, R.G., Sirker, J., Caux, J.-S., Hagemans, R.,
Maillet, J.M., White, S.R., Affleck, I.:
Dynamical structure factor at small $q$ for the XXZ
spin-1/2 chain.
J. Stat. Mech. P08022, 64 pp. (2007).\hfill\break
arXiv:0706.4327.

\end{thebibliography}
\end{document}